\documentstyle[12pt,epic]{article}
\newtheorem{theorem}{Theorem}[section]

\newtheorem{lemma}[theorem]{Lemma}

\renewcommand{\title}[1]{\large\bf \mbox{}\\ \mbox{}\\ \mbox{}\\ \mbox{}\\
     #1\bigskip\medskip\\}
\renewcommand{\author}[1]{\large #1\\ \smallskip}
\newcommand{\address}[1]{{\narrower\normalsize\it #1\\}\bigskip}
\newcommand{\zr}[1]{\mbox{\hspace*{#1em}}}
\newcommand{\hs}[1]{\hspace*{#1cm}}
\newcommand{\vs}[1]{\vspace*{#1cm}}
\newcommand{\sk}[1]{\vspace*{#1cm}\newline}
\newcommand{\Z}{\mbox{\sf Z\zr{-0.45}Z}}
\def\ade{$A$--$D$--$E$\space}
\def\wt#1#2#3#4#5#6{#1\!\!\mbox{
 $\left(\matrix{#5\vs{-0.1}&#4\cr#2&#3\cr}\biggm|\mbox{$#6$}\right)$}}
  
  \def\dddots{\mathinner{\mkern1mu\raise1pt\vbox{\kern1pt\hbox{.}}
                       \mkern2mu\raise4pt\hbox{.}
                       \mkern2mu\raise7pt\hbox{.}\mkern1mu}}
\def\wf#1#2#3#4#5#6#7#8#9{#1\mbox{$\left(
   \matrix{#5\vs{-0.1}&#8&#4\cr#9\vs{-0.1}&&#7\cr
a&#6&#3\cr}\biggm|\mbox{$#2$}\right)$}}

\def\row#1#2#3#4#5#6#7#8{#1\mbox{$\left(
   \matrix{#6&#8&#5\cr #3&#7&#4\cr}\biggm|\mbox{$#2$}\right)$}}
\def\Wf#1#2#3#4#5#6#7#8#9{W_{m\times n}\mbox{$\left(
   \matrix{#5&#8&#4\cr#9&&#7\cr #1&#6&#3\cr}\biggm|\mbox{$#2$}\right)$}}

\def\Wfml#1#2#3#4#5#6#7#8#9{W_{m\times l}\mbox{$\left(
   \matrix{#5&#8&#4\cr#9&&#7\cr #1&#6&#3\cr}\biggm|\mbox{$#2$}\right)$}}

\def\Wfln#1#2#3#4#5#6#7#8#9{W_{l\times n}\mbox{$\left(
   \matrix{#5&#8&#4\cr#9&&#7\cr #1&#6&#3\cr}\biggm|\mbox{$#2$}\right)$}}

\def\w11#1#2#3#4#5#6#7#8#9{W^{\!(1,1)}_{(1,1)}\mbox{$\left(
   \matrix{#5\vs{-0.1}&#8&#4\cr#9\vs{-0.1}&&#7\cr
#1&#6&#3\cr}\biggm|\mbox{$#2$}\right)$}}
\def\ba{\begin{array}}
\def\ea{\end{array}}
\def\be{\begin{eqnarray}}
\def\ee{\end{eqnarray}}
\def\no{\nonumber}
\def\m+{\!\!+\!\!}
\def\-{\!-\!}
\def\+{\!+\!}
\def\ol{\overline }
\def\scr{\scriptsize}
\def\for{\;\;{\rm for}\;}
\def\h{\hspace*{0.5cm}}
\def\mh{\hspace*{-0.5cm}}
\def\disp{\displaystyle}

\def\ra{\raisebox{0pt}[0pt][0pt]}
\def\ma{\makebox(0,0)[lb]}
\def\T{\mbox{\boldmath { $ T$}}}
\def\t{\mbox{\boldmath { $ t$}}}
\def\I{\mbox{\boldmath { $ I$}}}
\def\({\left(}
\def\){\right)}
\def\and{\hs{0.3}\mbox{and}\hs{0.3}}

\newcommand{\val}[1]{\mbox{val($#1$)}}

\def\face#1#2#3#4#5{\setlength{\unitlength}{0.0155in}%
  \begin{picture}(40,30)(-5,13)
  \thicklines
   \put(15,5){\makebox(0,0)[b]{\line(1,0){20}}}
   \put(5,5){\makebox(0,0)[b]{\line(0,1){20}}}
   \put(15,25){\makebox(0,0)[b]{\line(-1,0){20}}}
   \put(25,25){\makebox(0,0)[b]{\line(0,-1){20}}}
   \put(5,2){\makebox(0,0)[rb]{\scr $#1$}}  \put(26,2){\scr $#2$}
   \put(25,25){\scr $#3$} \put(5,25){\makebox(0,0)[rb]{\scr $#4$}}
   \put(13,13){\scr $#5$}
\end{picture}}

\def\YBR#1#2#3#4#5#6#7#8#9{
\begin{picture}(300,100)(-10,-10)
\thicklines
  \multiput(60,60)(0,-60){2}{\line(1,0){40}}
  \multiput(60,60)(60,-30){2}{\line(-2,-3){20}}
  \multiput(40,30)(60,30){2}{\line(2,-3){20}}
 \put(40,30){\line(1,0){40}}  \put(100,60){\line(-2,-3){20}}
 \put(100,0){\line(-2,3){20}} \put(80,30){\circle*{3}}
 \put(140,30){$=$}
  \multiput(190,60)(0,-60){2}{\line(1,0){40}}
  \multiput(190,60)(60,-30){2}{\line(-2,-3){20}}
  \multiput(170,30)(60,30){2}{\line(2,-3){20}}
 \put(190,60){\line(2,-3){20}}  \put(190,0){\line(2,3){20}}
 \put(250,30){\line(-1,0){40}}
 \put(210,30){\circle*{3}}
  \multiput(57,-7)(130,0){2}{\ma{\ra{\twlrm \scr $#1$}}}
  \multiput(100,-7)(130,0){2}{\ma{\ra{\twlrm \scr $#2$}}}
  \multiput(120,28)(130,0){2}{\ma{\ra{\twlrm \scr $#3$}}}
  \multiput(101,63)(130,0){2}{\ma{\ra{\twlrm \scr $#4$}}}
  \multiput(57,63)(130,0){2}{\ma{\ra{\twlrm \scr $#5$}}}
  \multiput(35,28)(130,0){2}{\ma{\ra{\twlrm \scr $#6$}}}
 \put(68,43){\ma{\ra{\twlrm \scr $#7$}}} 
 \put(95,28){\ma{\ra{\twlrm \scr $#8$}}} 
 \put(68,13){\ma{\ra{\twlrm \scr $#9$}}} 
 \put(218,43){\ma{\ra{\twlrm \scr $#9$}}} 
 \put(185,28){\ma{\ra{\twlrm \scr $#8$}}} 
 \put(218,13){\ma{\ra{\twlrm \scr $#7$}}} 
\end{picture}}
\def\diamond#1#2#3#4#5{\setlength{\unitlength}{0.0109in}%
\begin{picture}(66,48)(54,763)
\thicklines
 \put( 90,795){\line(-1,-1){ 30}}
 \put( 60,765){\line( 1,-1){ 30}}
 \put( 90,735){\line( 1, 1){ 30}}
 \put(120,765){\line(-1, 1){ 30}}
 \put( 72,777){\line(-1,-1){0}}
 \put(108,777){\line( 1,-1){0}}
 \put( 78,747){\line( 1,-1){0}}
 \put(102,747){\line(-1,-1){0}}
 \put( 56,762){\makebox(0,0)[rb]{\ra{\twlrm \scr $#1$}}}
 \put( 90,726){\ma{\ra{\twlrm \scr $#2$}}}
 \put(120,762){\ma{\ra{\twlrm \scr $#3$}}}
 \put(91,798){\makebox(0,0)[rb]{\ra{\twlrm \scr $#4$}}}
 \put( 87,762){\ma{\ra{\twlrm \scr $#5$}}}
\end{picture}}

\def\diagface#1#2#3#4#5#6#7{
\setlength{\unitlength}{.0943in}
\rule[-4.3\unitlength]{0in}{8.6\unitlength}
\begin{picture}(9,4)(-#6,-#7)
\put(1.5,0.5){\line(1,-1){3}}
\put(4.5,3.5){\line(1,-1){3}}
\put(4.5,3.5){\line(-1,-1){3}}
\put(7.5,0.5){\line(-1,-1){3}}
\put(1.2,0.5){\makebox(0,0)[r]{\small \mbox{$#1$}}}
\put(4.5,-2.8){\makebox(0,0)[t]{\small \mbox{$#2$}}}
\put(7.8,0.5){\makebox(0,0)[l]{\small \mbox{$#3$}}}
\put(4.5,3.8){\makebox(0,0)[b]{\small \mbox{$#4$}}}
\put(4.5,0.5){\makebox(0,0){\small \mbox{$#5$}}}
\end{picture}}
\def\1by2#1#2#3{\setlength{\unitlength}{0.0095in}%
\begin{picture}(108,110)(81,740)
\thicklines
\put( 93,768){\line( 1,-1){ 60}}
\put(153,708){\line( 1, 1){ 30}}
\put(183,738){\line(-1, 1){ 60}}
\put(123,798){\line(-1,-1){ 30}}
\put(153,768){\line(-1,-1){ 30}}
\put(135,735){\small$u\!+\!3\!\lambda$}
\put(118,766){\small$u$}
\put(60,770){\small$#1$}
\put(138,693){\small$#3$}
\put(188,730){\small$c$}
\put(158,765){\small$c'$}
\put(123,805){\small$d$}
\put(94,722){\small$#2$}
\end{picture}}

\def\dface#1#2#3#4#5{\setlength{\unitlength}{0.0090in}%
\begin{picture}(66,48)(54,763)
\thicklines
 \put( 90,795){\line(-1,-1){ 30}}
 \put( 60,765){\line( 1,-1){ 30}}
 \put( 90,735){\line( 1, 1){ 30}}
 \put(120,765){\line(-1, 1){ 30}}
 \put( 57,762){\makebox(0,0)[rb]{\ra{\twlrm \scr $#1$}}}
 \put( 90,726){\ma{\ra{\twlrm \scr $#2$}}}
 \put(120,762){\ma{\ra{\twlrm \scr $#3$}}}
 \put( 87,795){\ma{\ra{\twlrm \scr $#4$}}}
 \put( 98,762){\makebox(0,0)[rb]{\ra{\twlrm \scr $#5$}}}
\end{picture}}

\def\onebytwoface#1#2#3#4#5#6#7#8#9{
\begin{picture}(50,20)(-5,13)
\thicklines
 \put(9,13){\makebox(10,0)[b]{\scr $#7$}}
 \put(31,13){\makebox(10,0)[b]{\scr $#8$}}
 \multiput(5,5)(0,20){2}{\line(1,0){40}}
 \multiput(5,5)(20,0){3}{\line(0,1){20}}
 \multiput(25,5)(0.40000,-0.40000){9}{\makebox(0.4444,0.6667)#9}
 \multiput(25,5)(-0.40000,0.40000){9}{\makebox(0.4444,0.6667)#9}
 \multiput(25,5)(0.40000,0.40000){9}{\makebox(0.4444,0.6667)#9}
 \multiput(25,5)(-0.40000,-0.40000){9}{\makebox(0.4444,0.6667)#9}
 \put(3,2){\makebox(0,0)[rb]{\scr $#1$}}  \put(45,2){\scr $#2$}
 \put(45,26){\scr $#3$}  \put(3,26){\makebox(0,0)[rb]{\scr $#4$}}
 \put(21,-2){\makebox(10,0)[b]{\scr $#5$}}
 \put(21,28){\makebox(10,0)[b]{\scr $#6$}}   \put(48,14){}
\end{picture}}

\def\mnface{\begin{picture}(50,50)(-5,19)
 \put(23,23){\scriptsize u}
 \multiput(5,5)(0,40){2}{\line(1,0){40}}
 \multiput(5,5)(40,0){2}{\line(0,1){40}}
 \put(1,2){\scriptsize a} \put(24,-1){\scriptsize $\alpha$}
 \put(46,2){\scriptsize b}  \put(46,24){\scriptsize $\nu$}
 \put(46,46){\scriptsize c} \put(24,48){\scriptsize $\beta$}
 \put(0,46){\scriptsize d} \put(-2,24){\scriptsize $\mu$}
\end{picture}\h}
\def\mnfusion#1#2#3{ \put(171,657){\circle*{6}}
 \put(291,657){\circle*{6}}
 \put(336,657){\circle*{6}}
 \put(291,702){\circle*{6}}
 \put(336,747){\circle*{6}}
 \put(291,747){\circle*{6}}
 \put(171,702){\circle*{6}}
 \put(336,702){\circle*{6}}
 \put(171,747){\circle*{6}}
 \put(126,792){\line( 1, 0){255}}
 \put(381,792){\line( 0,-1){180}}
 \put(381,612){\line(-1, 0){255}}
 \put(126,612){\line( 0, 1){180}}
 \put(171,792){\line( 0,-1){ 45}}
 \put(171,702){\line( 0,-1){ 90}}
 \put(336,792){\line( 0,-1){ 45}}
 \put(336,702){\line( 0,-1){ 90}}
 \put(291,702){\line( 0,-1){ 90}}
 \put(291,792){\line( 0,-1){ 45}}
 \multiput(291,747)(0.00000,-8.18182){6}{\line( 0,-1){  4.091}}
 \multiput(336,747)(0.00000,-8.18182){6}{\line( 0,-1){  4.091}}
 \multiput(171,747)(0.00000,-8.18182){6}{\line( 0,-1){  4.091}}
 \multiput(171,747)(7.74194,0.00000){16}{\line( 1, 0){  3.871}}
 \multiput(171,702)(7.74194,0.00000){16}{\line( 1, 0){  3.871}}
 \multiput(171,657)(7.74194,0.00000){16}{\line( 1, 0){  3.871}}
 \put(126,747){\line( 1, 0){ 45}}
 \put(126,657){\line( 1, 0){ 45}}
 \put(126,702){\line( 1, 0){ 45}}
 \put(291,702){\line( 1, 0){ 90}}
 \put(291,657){\line( 1, 0){ 90}}
 \put(291,747){\line( 1, 0){ 90}}
 \put(292.5,675){\ma{\ra{\twlrm\tiny $u\!+\!(\!n\!\!-\!\! m\!)#1$}}}
 \put(291.5,630){\ma{\ra{\twlrm\tiny
$u\!\!+\!\!(\!\!n\!\!-\!\!m\!\!-\!\!1\!)#1$}}}
 \put(339,630){\ma{\ra{\twlrm\tiny $u\!\!+\!\!(\!n\!\!-\!\!m\!)#1$}}}
 \put(336.5,675){\ma{\ra{\twlrm\tiny
$u\!\!+\!\!(\!n\!\!\!-\!\!m\!\+\!1\!)#1$}}}
 \put(142,766){\ma{\ra {\twlrm \scr $u$}}}
 \put(127,628){\ma{\ra{\twlrm\tiny $u\!-\!(\!m\!- \!1)#1$}}}
 \put(127,675){\ma{\ra{\twlrm\tiny $u\!-\!(\!m\!-\!2)#1$}}}
 \put(339,766){\ma{\ra{\twlrm \tiny $u\!+\!(\!n\!- \!1)#1$}}}
 \put(293,767){\ma{\ra{\twlrm \tiny $u\!+\!(\!n\!-\! 2\!)#1$}}}
 \put(117,603){\ma{\ra{\twlrm \scr $a$}}}
 \put(117,795){\ma{\ra{\twlrm \scr $d$}}}
 \put(387,792){\ma{\ra{\twlrm \scr $c$}}}
 \put(387,603){\ma{\ra{\twlrm \scr $b$}}}
 \put(384,651){\ma{\ra{\twlrm \scr $\nu(b,c,m)_m$}}}
 \put(384,696){\ma{\ra{\twlrm \scr $\nu(b,c,m)_{m\!-\!1}$}}}
 \put(384,741){\ma{\ra{\twlrm \scr $\nu(b,c,m)_2$}}}
 \put(77,651){\ma{\ra{\twlrm \scr $q(\!d,\!a,\!m\!)_{\!j,m}$}}}
 \put(77,744){\ma{\ra{\twlrm \scr $q(d,a,m)_{j,2}$}}}
 \put(264,600){\ma{\ra{\twlrm \scr $p(\!a,\!b,\!n\!)_{\!i,n\!-\!1}$}}}
 \put(150,600){\ma{\ra{\twlrm \scr $p(a,b,n)_{i,2}$}}}
 \put(321,600){\ma{\ra{\twlrm \scr $p(a,b,n)_{i,n}$}}}
 \put(321,797){\ma{\ra{\twlrm \scr $\beta(d,c,n)_{n}$}}}
 \put(264,797){\ma{\ra{\twlrm \scr $\beta(d,c,n)_{n\!-\!1}$}}}
 \put(150,795){\ma{\ra{\twlrm \scr $\beta(d,c,n)_2$}}}
 \put(10,702){\ma{\ra{\twlrm\scr $\disp{\sum_{i,j}} \!
\phi^{(i,\alpha)}_{#3}(a,b)\phi^{(j,\mu)}_{#2}\!(a,d)$}}}
\end{picture}}

\def\up#1#2#3#4#5{
 \put( 74,675){\line( 1, 1){103.500}}
 \put(178,778){\line(-1, 1){ 26}}
 \put(152,804){\line(-1,-1){103.500}}
 \put( 48,701){\line( 1,-1){25.75}}
 \put(229,675){\line(-1, 1){103.500}}
 \put(152,804){\line( 1,-1){103}}
 \put(255,701){\line(-1,-1){ 25.75}}
 \put(100,752){\line( 1,-1){ 77.500}}
 \put(178,675){\line( 1, 1){ 51.500}}
 \put(203,752){\line(-1,-1){ 77.500}}
 \put(125,675){\line(-1, 1){ 51.500}}
 \put( 69,696){\ma{\ra{\twlrm \scr $-#1$}}}
 \put(170,696){\ma{\ra{\twlrm \scr $-#1$}}}
 \put(148,776){\ma{\ra{\twlrm \scr $u$}}}
 \put(107,750){\ma{\ra{\twlrm \scr $(1\-n)#1$}}}
 \put( 69,659){\ma{\ra{\twlrm \scr $#5$}}}
 \put(231,659){\ma{\ra{\twlrm \scr $#5$}}}
 \put(148,807){\ma{\ra{\twlrm \scr $a$}}}
 \put(120,780){\ma{\ra{\twlrm \scr $b_1$}}}
 \put( 36,699){\ma{\ra{\twlrm \scr $b_n$}}}
 \put(258,696){\ma{\ra{\twlrm \scr $#4$}}}
 \put(183,777){\ma{\ra{\twlrm \scr $#2$}}}
 \put(207,699){\ma{\ra{\twlrm \tiny $u\+(\!n\-1\!)\!#1$}}}
 \put(207,753){\ma{\ra{\twlrm \scr $#3$}}}
\end{picture}}
\def\down#1{
 \put( 86,786){\line( 1,-1){103.500}}
 \put(164,657){\line(-1, 1){103.500}}
 \put( 60,760){\line( 1, 1){25.75}}
 \put(241,786){\line(-1,-1){103.500}}
 \put(164,657){\line( 1, 1){103}}
 \put(267,760){\line(-1, 1){25.75}}
 \put(112,709){\line( 1, 1){ 77.500}}
 \put(190,786){\line( 1,-1){ 51.500}}
 \put(215,709){\line(-1, 1){ 77.500}}
 \put(137,786){\line(-1,-1){ 51.500}}
 \put(160,685){\ma{\ra{\twlrm \scr $u$}}}
 \put(161,650){\ma{\ra{\twlrm \scr $b$}}}
 \put( 84,793){\ma{\ra{\twlrm \scr $a$}}}
 \put(243,793){\ma{\ra{\twlrm \scr $a$}}}
 \put(230,759){\ma{\ra{\twlrm \scr $-#1$}}}
 \put( 65,759){\ma{\ra{\twlrm \tiny $u\+(\!n\-1\!)\!#1$}}}
 \put(134,762){\ma{\ra{\twlrm \scr $-#1$}}}
 \put(169,706){\ma{\ra{\twlrm \scr $(1\-n)#1$}}}
 \put(195,675){\ma{\ra{\twlrm \scr $a_n$}}}
 \put(126,672){\ma{\ra{\twlrm \scr $b_n$}}}
 \put( 48,756){\ma{\ra{\twlrm \scr $b_1$}}}
 \put(270,759){\ma{\ra{\twlrm \scr $a_1$}}}
\end{picture}}
\def\upup#1{
 \put( 74,675){\line( 1, 1){103.500}}
 \put(152,804){\line(-1,-1){103.500}}
 \put( 48,701){\line( 1,-1){25.75}}
 \put(229,675){\line(-1, 1){103.500}}
 \put(152,804){\line( 1,-1){103}}
 \put(255,701){\line(-1,-1){25.75}}
 \put(203,752){\line(-1,-1){ 77.500}}
 \put(125,675){\line(-1, 1){ 51.500}}
 \put(100,752){\line( 1,-1){ 77.500}}
 \put(178,675){\line( 1, 1){ 51.500}}
 \put(113,699){\ma{\ra{\twlrm \scr $-#1$}}}
 \put( 70,698){\ma{\ra{\twlrm \scr $u$}}}
 \put(181,780){\ma{\ra{\twlrm \scr $a$}}}
 \put(153,804){\ma{\ra{\twlrm \scr $b_1$}}}
 \put( 41,698){\ma{\ra{\twlrm \scr $b$}}}
 \put( 69,663){\ma{\ra{\twlrm \scr $a_n$}}}
 \put(258,699){\ma{\ra{\twlrm \scr $a_{n-1}$}}}
 \put(228,663){\ma{\ra{\twlrm \scr $a_n$}}}
 \put(218,699){\ma{\ra{\twlrm \scr $-#1$}}}
 \put(132,776){\ma{\ra{\twlrm \tiny $u\!\!+\!\!(n\!\!-\!\!1\!)\!#1$}}}
 \put(158,750){\ma{\ra{\twlrm \scr $(1\-n)#1$}}}
\end{picture}}

\def\Poperator#1{
 \put(210,780){\line(-1,-1){150}}
 \put(210,780){\line( 1,-1){150}}
 \put(210,780){\line(-1,-1){150}}
 \put( 60,630){\line( 1,-1){ 30}}
 \put( 90,600){\line( 1, 1){150}}
 \put(120,690){\line( 1,-1){ 90}}
 \put(210,600){\line( 1, 1){ 90}}
 \put(150,720){\line( 1,-1){120}}
 \put(270,600){\line( 1, 1){ 60}}
 \put(180,750){\line( 1,-1){150}}
 \put(330,600){\line( 1, 1){ 30}}
 \put( 90,660){\line( 1,-1){ 60}}
 \put(150,600){\line( 1, 1){120}}
 \put( 47,627){\ma{\ra{\twlrm \scr $b_n$}}}
 \put( 90,581){\ma{\ra{\twlrm \scr $b$}}}
 \put(330,584){\ma{\ra{\twlrm \scr $b$}}}
 \put(208,786){\ma{\ra{\twlrm \scr $a$}}}
 \put(170,755){\ma{\ra{\twlrm \scr $b_1$}}}
 \put(141,724){\ma{\ra{\twlrm \scr $b_2$}}}
 \put(239,755){\ma{\ra{\twlrm \scr $a_1$}}}
 \put(266,724){\ma{\ra{\twlrm \scr $a_2$}}}
 \put(306,626){\ma{\ra{\twlrm \tiny $u\!\!+\!\!(n\!\!-\!\!1)#1$}}}
 \put(363,630){\ma{\ra{\twlrm \scr $a_n$}}}
 \put(208,749){\ma{\ra{\twlrm \scr $u$}}}
 \put(161,717){\ma{\ra{\twlrm \scr $(1\!\!-\!\!n)#1$}}}
 \put(226,717){\ma{\ra{\twlrm \scr $u\+#1$}}}
 \put(227,653){\ma{\ra{\twlrm \scr $-2#1$}}}
 \put(110,654){\ma{\ra{\twlrm \scr $-2#1$}}}
 \put( 89,622){\ma{\ra{\twlrm \scr $-#1$}}}
 \put(264,625){\ma{\ra{\twlrm \scr $-#1$}}}
\end{picture}}
\def\pushthroughone#1#2#3#4#5{ \put(201,753){\line( 0,-1){ 30}}
 \put(231,753){\line( 0,-1){ 30}}
 \put(171,753){\line( 0,-1){ 30}}
 \put(111,753){\line( 0,-1){ 30}}
 \multiput(171,753)(30,0){3}{\circle*{5}}
 \put(111,723){\line( 1, 0){195}}
 \put(306,723){\line( 0, 1){ 30}}
 \put(306,753){\line(-1, 0){195}}
 \put(309,711){\ma{\ra{\twlrm \scr $b$}}}
 \put(210,732){\ma{\ra{\twlrm \scr $u$}}}
 \put(173,732){\ma{\ra{\twlrm \scr $u-#1$}}}
 \put(185,711){\ma{\ra{\twlrm \tiny $p(a,\!b,\!#2)_{j\!,\!i}$}}}
 \put(183,711){\makebox(0,0)[rb]{\ra{\twlrm \tiny $p(a,\!b,\!#2)_{j\!,i\-1}$}}}
 \put(#4,711){\makebox(0,0)[lb]{\ra{\twlrm \tiny
$p(a,\!b,\!#2)_{\!j\!,i\+1}$}}}
 \put(111,783){\line( 0,-1){ 30}}
 \put(111,753){\line( 1, 0){195}}
 \put(306,753){\line( 0, 1){ 30}}
 \put(306,783){\line(-1, 0){195}}
 \put(171,783){\line( 0,-1){ 30}}
 \put(201,783){\line( 0,-1){ 30}}
 \put(231,783){\line( 0,-1){ 30}}
 \put(165,786){\ma{\ra{\twlrm \scr $c_{i\-1}$}}}
 \put(198,786){\ma{\ra{\twlrm \scr $c_i$}}}
 \put(222,786){\ma{\ra{\twlrm \scr $c_{i\+1}$}}}
 \put(173,762){\ma{\ra{\twlrm \scr $v-#1$}}}
 \put(210,762){\ma{\ra{\twlrm \scr $v$}}}
 \put(312,783){\ma{\ra{\twlrm \scr $c$}}}
 \put(102,786){\ma{\ra{\twlrm \scr $d$}}}
 \put(32,747){\ma{\ra{\twlrm \scr $\sum_j^{P_{(a,\!b)}^{#5}}
\!\!\!\!\!\!\phi^{(j,\alpha)}_{#3}\!\!(\!a\!,\!b\!)$}}}
 \put(104,747){\ma{\ra{\twlrm \scr $e$}}}
 \put(312,747){\ma{\ra{\twlrm \scr $f$}}}
 \put(105,714){\ma{\ra{\twlrm \scr $a$}}}
\end{picture}}
\def\pushthroughtwo#1#2#3#4#5{ \put(216,717){\line( 0,-1){ 30}}
 \put(246,717){\line( 0,-1){ 30}}
 \put(186,717){\line( 0,-1){ 30}}
 \put(126,717){\line( 0,-1){ 30}}
 \put(126,687){\line( 1, 0){195}}
 \put(321,687){\line( 0, 1){ 30}}
 \put(321,717){\line(-1, 0){195}}
 \put(213,783){\line( 0,-1){ 30}}
 \put(243,783){\line( 0,-1){ 30}}
 \put(183,783){\line( 0,-1){ 30}}
 \put(123,783){\line( 0,-1){ 30}}
 \put(123,753){\line( 1, 0){195}}
 \put(318,753){\line( 0, 1){ 30}}
 \put(318,783){\line(-1, 0){195}}
 \put(324,675){\ma{\ra{\twlrm \scr $b$}}}
 \put(227,696){\ma{\ra{\twlrm \scr $u$}}}
 \put(189,696){\ma{\ra{\twlrm \scr $u\-#1$}}}
 \put(197,675){\ma{\ra{\twlrm \tiny $p(a,\!b,\!#2)_{j,i}$}}}
 \put(195,675){\makebox(0,0)[rb]{\ra{\twlrm \tiny $p(a,\!b,\!#2)_{j,i\-1}$}}}
 \put(#4,675){\ma{\ra{\twlrm \tiny $p(a,\!b,\!#2)_{j,i\+1}$}}}
 \put(197,720){\ma{\ra{\twlrm \tiny $\beta(e,\!f,\!#2)_i$}}}
 \put(196,720){\makebox(0,0)[rb]{\ra{\twlrm \tiny $\beta(e,\!f,\!#2)_{i\-1}$}}}
 \put(#4,720){\ma{\ra{\twlrm \tiny $\beta(e,\!f,\!#2)_{i\+1}$}}}
 \put(117,720){\ma{\ra{\twlrm \scr $e$}}}
 \put(114,744){\ma{\ra{\twlrm \scr $e$}}}
 \put(327,714){\ma{\ra{\twlrm \scr $f$}}}
 \put(324,744){\ma{\ra{\twlrm \scr $f$}}}
 \put(195,741){\makebox(0,0)[rb]{\ra{\twlrm \tiny $q(e,\!f,\!#2)_{k,i\-1}$}}}
 \put(197,741){\ma{\ra{\twlrm \tiny $q(e,\!f,\!#2)_{k,i}$}}}
 \put(#4,741){\ma{\ra{\twlrm \tiny $q(e,\!f,\!#2)_{k,i\+1}$}}}
 \put(189,762){\ma{\ra{\twlrm \scr $v\-#1$}}}
 \put(227,762){\ma{\ra{\twlrm \scr $v$}}}
 \put(210,786){\ma{\ra{\twlrm \scr $c_i$}}}
 \put(240,786){\ma{\ra{\twlrm \scr $c_{i\+1}$}}}
 \put(177,786){\ma{\ra{\twlrm \scr $c_{i\-1}$}}}
 \put(324,783){\ma{\ra{\twlrm \scr $c$}}}
 \put(114,783){\ma{\ra{\twlrm \scr $d$}}}
 \put(114,681){\ma{\ra{\twlrm \scr $a$}}}
 \put( 50,762){\ma{\ra{\twlrm \scr $\sum_k^{P_{(e,\!f)}^{#5}}
            \!\!\!\!\!\!\phi^{(\!k\!,\beta\!)}_{#3}\!\!(\!e\!,\!f\!)$}}}
 \put( 53,699){\ma{\ra{\twlrm \scr $\sum_j^{P_{(a,\!b)}^{#5}}
            \!\!\!\!\!\phi^{(\!j\!,\alpha\!)}_{#3}\!\!(\!a\!,\!b\!)$}}}
 \put(48,729){\ma{\ra{\twlrm \sc$\sum_{\beta=1}^{A^{#3}_{(e,f)}}$}}}
\end{picture}}

\begin{document}

\newlength{\origbaselineskip}
\setlength{\origbaselineskip}{\baselineskip}

\vspace{0.5cm}
\begin{center}
\title{{\boldmath $SU$}(2) HIERARCHIES  OF DILUTE \\ LATTICE MODELS }
\author{\sc Yu-kui Zhou\footnote{Email: ykzhou@mundoe.maths.mu.oz.au} }
\address{\em Mathematics Department, University of Melbourne,\\
            Parkville, Victoria 3052, Australia }

\begin{abstract}
\begin{quotation}
{ The fusion procedure of dilute $A_L$ models is constructed.
It has been shown that the fusion rules have two types: $su(2)$
and $su(3)$.  This paper is concerned with the $su(2)$ fusion rule
mainly and the corresponding functional relations of commuting transfer
matrices in the $su(2)$ fusion hierarchy are found. Specially, it
has been found that the fusion hierarchy does not close.
These two types of fusion generate different solvable models, but, they
are not totally irrelevant. The $su(2)$ fusion of level $2$ is equivalent
to the $su(3)$ fusion of level $(1,1)$. According to this relationship
the Bethe ansatz of fused model of level $(1,1)$ in $su(3)$
hierarchy has been represented by that of level $2$ in $su(2)$ fusion
hierarchy. }
 \end{quotation}
\end{abstract}
\end{center}

\section{Introduction}
In the family of restricted solid--on--solid (RSOS) models
the dilute $A_L$ lattice models \cite{WNS:92} are very new
two-dimensional solvable models and are the generalization of
 the ABF's RSOS models \cite{ABF:84}. At criticality
the dilute $A_L$ lattice models \cite{Roch:92} admit the $D$ or $E$
extension like the critical \ade models \cite{Pasquier:87}.
Thus the intertwiners have been constructed among these dilute
\ade models \cite{PeZh:93}.

The dilute $A_L$ lattice models are built on the $A_L$ Dynkin diagram
with a loop at each spin node as shown in Figure~1. The face weights
have the crossing symmetry and satisfy the inversion relations. The models
at off criticality do not satisfy the $\Z_2$ symmetry\footnote{In fact,
the $\Z_2$ symmetry is broken only for odd $L$.} of the adjacency condition
graph Figure~1. Recent studies have shown that the dilute $A_L$ models
present many interesting aspects. In an appropriate regime the dilute
$A_3$ model lies in the universality class of the Ising model in a
magnetic field and gives the magnetic exponent $\delta=15$ \cite{WPSN:94}.
Also the $A_3$ model shows the $E_8$ scattering theory for massive
excitations over the ground state \cite{BNW:94,Zam:89}.
\begin{figure}[t]
\begin{center}
\setlength{\unitlength}{0.0125in}%
\begin{picture}(278,42)(49,732)
\thicklines
\put(61,762){\circle{24}}
\put(105,762){\circle{24}}
\put(150,762){\circle{24}}
\put(315,762){\circle{24}}
\put(270,762){\circle{24}}
\put(60,750){\circle*{6}}
\put(105,750){\circle*{6}}
\put(150,750){\circle*{6}}
\put(270,750){\circle*{6}}
\put(315,750){\circle*{6}}
\put(60,750){\line(1,0){255}}
\put(56,737){\small$1$}
\put(103,737){\small$2$}
\put(148,737){\small$3$}
\put(311,739){\scr$L$}
\put(260,739){\scr$L\!-\!1$}
\end{picture}
\caption{The graph of adjacency condition of the dilute $A_L$ models.}
\end{center}
\end{figure}
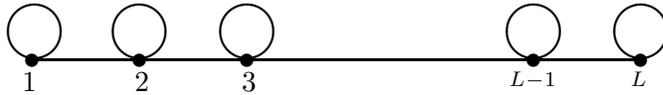

Fusion procedure is very useful in studying two-dimensional solvable
vertex and face models
\cite{KRS:81,DJKMO:86,DJKMO:88,JKMO:88b,ZhHo:89,ZhPe:94}. Essentially,
fusion enables the construction of new solutions to the Yang-Baxter equations
or star-triangle relation
\cite{Yang:67,Baxter:82} from a given fundamental solution.
The fusion procedure of the dilute models has not been constructed.
In this paper we construct the fusion of the dilute A models. This is completed
by the method expressed in \cite{ZhPe:94}. It can be shown that the models
admit
two types of fusion. The first one is $su(2)$-type and the other one is
$su(3)$-type fusion. They are very different and employ different projectors
for constructing  fused face weights. We present the $su(2)$-fusion only in
this
paper. Differently the $su(3)$-fusion is more complicated and will be published
elsewhere \cite{ZPG:94}.

 In next section we describe the dilute lattice models.
The face weights satisfy a group of  special properties which ensure that
they can be taken as the elementary blocks for fusion. It is shown that
there are two projectors, which generate two different types of fusion:
$su(2)$ and $su(3)$ fusion based on the elementary models. In section~3
we give in detail
the procedure for constructing the level $2$ fused face weights of the
$su(2)$-type. Then we construct the general procedure for finding the
the $su(2)$ fused face weights of level $n$. This is accomplished by
introducing "coordinates" on the independent paths of the fusion projectors.
In section~4 we discuss the relationship between the $su(2)$ fusion and $su(3)$
fusion. It is shown that the fused model of level $(1,1)$ in
the $su(3)$ fusion hierarchy is
equivalent to the one of level $2$ in the $su(2)$ fusion hierarchy.
In section~5 we describe the $su(2)$ functional relations satisfied by the
fused dilute $A_L$ row transfer matrices. In section~6  the eigenvalues of
transfer matrices of the $su(2)$ fusion hierarchy are discussed.
In particular, the Bethe ansatz is also presented for the fused $(1,1)$ model
of
the $su(3)$ fusion hierarchy according to the relationship between these
two fusions. In the final
section a brief discussion is presented.
\section{ Elementary block}
\setcounter{equation}{0}
In this section we express the face weights of the dilute $A_L$ models and
list the properties of the face weighs. These properties are useful for
constructing the fusion.

The states at adjacent sites of the dilute $A_L$ square lattice must be
adjacent
on the graph in Figure~1. The face weights  of the models  not satisfying
this adjacency condition for each pair of adjacent sites around a face vanish.
The nonzero face weights of the dilute $A_L$ models (off-critical)
are given by \cite{WNS:92}
\be
&&\wt W{\;a}{a\;}{a\;}{\;a}u
\;\;=\;{\vartheta_1(6\lambda-u)\vartheta_1(3\lambda+u)\over
              \vartheta_1(6\lambda)\vartheta_1(3\lambda)}-\({S(a+1)\over S(a)}
              {\vartheta_4(2a\lambda-5\lambda)\over
              \vartheta_4(2a\lambda+\lambda)}\right.\no\\
     &&\hs{2.5}+\left.{S(a-1)\over S(a)}{\vartheta_4(2a\lambda+5\lambda)\over
       \vartheta_4(2a\lambda-\lambda)}\){\vartheta_1(u)\vartheta_1(3\lambda-u)
       \over\vartheta_1(6\lambda)\vartheta_1(3\lambda)} \no\\ \no\\
&&\wt Wa{a\pm 1}{a\pm 1}au = \wt Waa{a\pm 1}{a\pm 1}u =
        {\vartheta_1(u)\vartheta_1(3\lambda-u)\over
        \vartheta_1(2\lambda)\vartheta_1(3\lambda)}  \no\\
&&\hs{2.5} \({\vartheta_4(\pm 2a\lambda+3\lambda)\vartheta_4(\pm
2a\lambda-\lambda)
     \over\vartheta_4^2(\pm 2a\lambda+\lambda)}\)^{1/2}
       \no\\ \no\\
&&\wt Waaa{a\pm 1}u \;=\; \wt Wa{a\pm 1}aau =
         {\vartheta_1(3\lambda-u)\vartheta_4(\pm 2a\lambda+\lambda-u)\over
          \vartheta_1(3\lambda)\vartheta_4(\pm 2a\lambda+\lambda)}\no\\ \no\\
&&\wt W{a\pm 1}aaau \;=\; \wt Waa{a\pm 1}au =\({S(a\pm 1)\over S(a)}\)^{1/2}
        {\vartheta_1(u)\vartheta_4(\pm 2a\lambda-2\lambda+u)\over
        \vartheta_1(3\lambda)\vartheta_4(\pm 2a\lambda+\lambda)} \no\\ \no\\
&&\wt Wa{a\pm 1}a{a\mp 1}u \;=\;
              {\vartheta_1(2\lambda-u)\vartheta_1(3\lambda-u)\over
              \vartheta_1(2\lambda)\vartheta_1(3\lambda)}  \label{eBlock}\\
\no\\
&&\wt W{a\pm 1}a{a\mp 1}au \;=\;-\({S(a-1)S(a+1)\over S^2(a)}\)^{1/2}
             {\vartheta_1(u)\vartheta_1(\lambda-u)\over
              \vartheta_1(2\lambda)\vartheta_1(3\lambda)} \no\\ \no\\
&&\wt W{a\pm 1}a{a\pm 1}au \;=\;
        {\vartheta_1(3\lambda-u)\vartheta_1(\pm 4a\lambda+2\lambda+u)\over
          \vartheta_1(3\lambda)\vartheta_1(\pm 4a\lambda+2\lambda)} \no\\
         &&\hs{2.5} +\;{S(a\pm1)\over S(a)}
        {\vartheta_1(u)\vartheta_1(\pm 4a\lambda-\lambda+u)\over
          \vartheta_1(3\lambda)\vartheta_1(\pm 4a\lambda+2\lambda)} \no
\ee\smallskip
where the spectral parameter is $u$ and the crossing parameter
$\lambda=\frac{\pi}{4}\frac{L+2}{L+1}$ or $
\frac{\pi}{4}\frac{L}{L+1}$. These $S(a)$ are given by
$$
S(a)  =  (-1)^{\displaystyle a} \;
\frac{\vartheta_1(4a\lambda)}{\vartheta_4(2a\lambda)} \, . $$
The elliptic functions $\vartheta_1(u),\vartheta_4(u)$ are the Jacobian
$\vartheta$-functions
of $nome$ $p$,
\be
\vartheta_1(u)&=&2p^{1/4}\sin u\prod_{n=1}^\infty
    (1-2p^{2n}\cos 2u+p^{4n})(1-p^{2n}) \no \\
\vartheta_4(u)&=&\prod_{n=1}^\infty
    (1-2p^{2n-1}\cos 2u+p^{4n-2})(1-p^{2n}) \no
\ee
In critical limit $p\to 0$ the function $\vartheta_1(u)/\vartheta_1(\lambda)
\sim \sin u/\sin\lambda$ and $\vartheta_4(u)\sim 1$. Under this limit
 these $S(a)$ are reduced
to the elements $S_a$ of the Perron-Frobenius eigenvectors $S$ of
the adjacency matrix $A$ defined by the classical $A_L$ Dynkin diagram
with the elements
\begin{equation}
A_{a,b}=\left\{ \ba{ll}
             1,\qquad & \mbox{$|a-b|=1$ } \\
             0,\qquad & \mbox{otherwise.}
                 \ea  \right.
\label{ABFadj}\end{equation}

The nonzero face weights can be graphically represented by the four
spins surrounding square faces, or the graphs given by  rotating clockwise
these square faces through $\pi\over 2$. The face weights satisfy the
following crossing \vs{-0.3} symmetry
\be
\face abcd{\hs{-.27}3\!\lambda\!-\!u} =\;\;\;
   \Biggr({S(a)S(c)\over S(b)S(d)}\Biggr)^{1/2}
   \face dabcu  \h ,
\ee
the inversion relation of the form
 \vs{0.2}\be
&&\hs{1.}\sum_{c}
\begin{picture}(132,30)(54,763)
\put(68,763){\diamond {a}b{}du}
\put(114,763){\diamond {}b{a'}d{\-u}}
\put(117,758){\scr $c$}\end{picture} \no \\ \no \\ \no \\
&&=
\disp{\vartheta_1(2\lambda+u)\vartheta_1(3\lambda+u)
 \vartheta_1(2\lambda-u)\vartheta_1(3\lambda-u)\over
\vartheta_1^2(2\lambda)\vartheta_1^2(3\lambda)}
\label{eq:inversion}
\ee
and the star-triangle relation
\be
 \YBR abcdefu{u\-v}v \label{eq:YBR}
\ee
where the sum is meant by the solid circle. The inversion relation vanishes for
$u=\lambda_2:=2\lambda$ or $u=\lambda_1:=3\lambda$. This follows that the
following two pairs of operators
$$ \mbox{$\dface {}b{}d{\lambda_1}\;$ , $\dface {}b{}d{\-\lambda_1}\h$ and }
\mbox{$\h\dface {}b{}d{\lambda_2}\;$ , $\dface {}b{}d{\-\lambda_2}$} $$
\sk{0.15}
are singular. Each pair of them is orthogonal and can be taken as the
projectors of fusion. So two different types of fused models can be constructed
from the elementary dilute models (\ref{eBlock}).

The projectors for the first fusion group are the face weights with the
spectral parameters $\pm\lambda_1$ and they acquire the following
properties
\be
&&\hs{2.}\dface a{\hs{-0.3}b\ne d}cd{\lambda_1\!\!} \hs{0.2}=0
  \; ,\hs{1}\label{zero}\\&&\no\\&&\no\\
&&\dface {b\!\pm\! 1}bcb{\lambda_1\!\!} \hs{0.2}
     =\; A_{b\!\pm\! 1,b} \Biggr({S(b\pm 1)\over S(b)}\Biggr)^{1/2}\hs{0.2}
                             \dface bbcb{\lambda_1\!\!}
                         \label{oneg1p1}  \\&&\no
\ee
and inserting (\ref{oneg1p1}) into the inversion
relation (\ref{eq:inversion}) we have
\be
&&\hs{-1.4}\dface ab{b}b{\-\lambda_1\!\!} \hs{0.2} =-
\Biggr({S(b\- 1)\over S(b)}\Biggr)^{1/2}\dface ab{b\-1}b{\-\lambda_1\!\!}
\h - \Biggr({\;S(b\+ 1)\over S(b)}\Biggr)^{1/2}
 \dface ab{b\+1}b{\-\lambda_1\!\!}  \h\h .
                           \label{oneg2p1}        \\&&\no
\ee
For the second fusion group the projectors satisfy the following
properties
\be
&&\hs{2.7}\dface a{\hs{-0.2}b\!\pm\! 2}ab{\lambda_2\!\!}
                  \hs{0.2}=0\\&&\no\\&&\no\\
&&\dface {b\!\pm\! 1}bcb{\lambda_2\!\!} \hs{0.15}=A_{b\!\pm\! 1,b}\;
    \disp{\vartheta_1(\lambda)\over\vartheta_1(2\lambda)}
    \Biggr(\disp{S(a\pm 1,a)\over S(a,a)}\hs{0.14}\Biggr)^{1/2}
    \dface bbcb{\lambda_2\!\!}\hs{0.2}
                 \label{twog1p1}\\&&\no\\&&\no\\
&&\hs{1.1}\dface bbc{\!\!b\+1}{\lambda_2} \h =\;
       \hs{0.7} \dface {\!b\+1}bc{\!\!b\+1}{\lambda_2}
                 \label{twog1p2}\\&&\no
\ee
and inserting (\ref{twog1p1}) and (\ref{twog1p2}) into the inversion
relation (\ref{eq:inversion}) we have
\be
&&\hs{-1.9}\Biggr({S(b\- 1,b)\over S(b,b)}\Biggr)^{1/2}\!
        \dface ab{b\-1}b{\-\lambda_2\!\!}
    \hs{0.6}+\Biggr({S(b\+ 1,b)\over S(b,b)}\Biggr)^{1/2}
        \!\dface ab{b\+1}b{\-\lambda_2\!\!} \h =
     -{\vartheta_1(2\lambda)\over\vartheta_1(\lambda)}
        \;\dface abbb{\-\lambda_2\!\!} \hs{0.2}
                              \label{twog2p1}    \\&&\no\\&&\no\\
&&\hs{3.2}\dface abb{\!\!b\pm 1}{\-\lambda_2} \h=\;-\;
  \dface ab{b\pm 1}{\!\!b\pm 1}{\-\lambda_2} \label{twog2p2} \\&&\no
\ee
where
\be
S(a,b)&=& S(a) f(a,b) \\
f(a,b)&=& \({\vartheta_4[2(a-b)b\lambda-\lambda]\over
    \vartheta_4[2(a-b)b\lambda]}\)^{2|a-b|} \h .\label{f}
\ee
The properties of the projectors for the critical dilute
models can be given by taking critical limit $p\to 0$ in
(\ref{zero})--(\ref{f}).

The adjacency condition of the dilute $A_L$ models can be represented
by the classical $A_L$ Dynkin diagram with a loop at each node. Each
node $a$ of the diagram has a coordination or valence $\val a=2,3$,
or
\begin{equation}
\val a=\sum_b (\delta_{a,b}+A_{a,b})\;.
\end{equation}
The number of nonzero terms in (\ref{oneg2p1}) and (\ref{twog2p1}) is given
exactly by $\val b$. Specifically, we have the valence $\val b=3$ for most of
the
node $b$ except for the endpoints $1$ and $L$,
which have $\val 1 =\val L =2$. For the ABF model \cite{ABF:84} it has the
classical $A_L$ Dynkin diagram as the adjacency condition and the valence
is less than $3$: $1$ for the endpoints $1,L$ and $2$ for the other nodes.
So the fusion
for the dilute models is more complicated and proceeds differently.

The fusion for the second group has been given in \cite{ZPG:94}.
In the following sections we will show that the first group fusion is of
$su(2)$ type and the second one is of $su(3)$ type.
We will describe the $su(2)$ fusion procedure in detail
and also discuss the relation between these two fusions.

\section{Fusion for shift $\lambda_1$}
\setcounter{equation}{0}

\subsection{Admissibility}
\setcounter{equation}{0}

The adjacency matrix $A$ of the models (\ref{eBlock}) satisfies the following
fusion rule,
\begin{eqnarray}
& & A^{(n)} A^{(1)}=A^{(n-1)}+A^{(n+1)}\;,n=1,2,\cdots \no \\
& & A^{(0)}={\bf I},\; \;A^{(1)}={\bf I}+A\;\; \label{eq:adjfusionone}
\end{eqnarray}
where ${\bf I}$ is the unitary matrix and $A$ is the adjacency matrix defined
in
(\ref{ABFadj}). The fusion rule
(\ref{eq:adjfusionone}) in form likes the one of  the classical \ade models
and thus this fusion is of $su(2)$ type. From (\ref{eq:adjfusionone}) we can
obtain
\be
A^{(n)} A^{(n)}=A^{(n-1)}A^{(n+1)}+\mbox{\bf I }\; .\label{AA}
\ee

The matrices $A^{(n)}$ are the adjacency matrices of the fused face weights
built using the projector with the spectral parameter shift $\lambda_1$.
Unlike the classical \ade models the fusion of the dilute models with
the shift $\lambda_1$ can go any higher fusion levels
and hence do not have the closure condition.
The elements of $A^{(n)}$ can in general be nonnegative integers greater
than one. In this case we distinguish the edges of the adjacency diagram
joining two given sites by bond variables $\alpha =1,2,\cdots$ If there
is just one edge then the corresponding bond variable is $\alpha=1$.

These fusion rules (\ref{eq:adjfusionone}) and (\ref{AA}) can be extended
to the level of transfer matrices, which give the functional relations
for the transfer matrices and are constructed by the fusion procedure.

\subsection{1 by 2 Fusion}
We implement the elementary fusion of one by two block of face weights,
which  are the  symmetric $1\times 2$ fusion using the projector
appeared in (\ref{oneg2p1}) and the antisymmetric $1\times 2$ fusion
using the projectors appeared
in (\ref{oneg1p1}). Notice that in the level 2 fused models, the
occurrence of bond variables on the edges of the symmetric fused
face weights arises when both adjacent sites are the same spin with
valence $\val a=3$ or differ by $1$ on the spin.
The antisymmetric $1\times 2$ fusion gives a
trivial solution of the YBR because the nonzero fused face weights are
all equal.

The symmetric fusion of level $2$ can be expressed by the following
objector.
\be
\setlength{\unitlength}{0.0107in}%
\begin{picture}(135,117)(153,660)
\thicklines
\put(195,735){\line(-1,-1){ 30}}
\put(165,705){\line( 1,-1){ 30}}
\put(195,675){\line( 1, 1){ 60}}
\put(255,735){\line(-1, 1){ 30}}
\put(225,765){\line(-1,-1){ 30}}
\put(195,735){\line( 1,-1){ 60}}
\put(255,675){\line( 1, 1){ 30}}
\put(285,705){\line(-1, 1){ 30}}
\put(182,702){\scr$\!-3\!\lambda$}
\put(222,732){\scr$u$}
\put(241,702){\scr$u\!+\!3\!\lambda$}
\put(258,735){\scr$c'$}
\put(192,660){\scr$b$}
\put(252,660){\scr$b$}
\put(153,702){\scr$b'$}
\put(186,738){\scr$a$}
\put(225,771){\scr$d$}
\put(288,702){\scr$c$}
\put(225,705){\circle*{4}}
\end{picture} \label{12fusion}
\ee
where the solid circle means sum over all possible spins, say,
sum over $a'$.
However, the summation vanishes if $a'$ is not admissible to
the neighbor spins $a$ and $b$ in the adjacency graph Figure~1.
Thus using the property (\ref{oneg2p1}) the objector (\ref{12fusion})
can be expressed
as the following cases.
\newline
1) \vs{-1.6} $a\ne b$:
\be
\sum_{a'}\;\;\thicklines\diagface {b'}b{a'}{a}{-3\!\lambda}00
\1by2 {\h a}{\hs{0.4}a'}{\hs{0.2} b}  \label{12-1}
\ee
where $a'=a$ or $b$ for $|a-b|=1$ and $a'=(a+b)/2$ for $|a-b|=2$.
\newline 2) \vs{-1} $a=b$ and $\val b =2$:
\be
\thicklines\diagface {b'}b{b}{b}{-3\!\lambda}00
 \h\Biggr(\1by2 {\hs{0.6}b}{\h b}{\hs{0.2} b}\h
 \mbox{$-\Biggr(\disp{S(b)\over S(b_1)}\Biggr)^{1/2}$}
   \h\1by2 {\hs{0.6}b}{\h b_1}{\hs{0.2} b}\h \Biggr)  \label{12-2}
\ee
where $b_1=2$ for $b=1$ and $b_1=L-1$ for $b=L$.
\newline 3) \vs{-1} $a=b$ and $\val b =3$:
\be
&\hs{-0.5}\thicklines\diagface {b'}{b}{b\!+\!1}{b}{-3\!\lambda}00
\h\Biggr( \hs{0.2}\1by2 {\h b}{b\!+\!1}{\h b} \h-
  \Biggr(\disp{S(b\!+\!1)\over S(b)}\Biggr)^{1/2}
    \h\1by2 {\h b}{\h b}{\h b} \hs{0.3}\Biggr) & \no\\
 &+\;\;\thicklines\diagface {b'}{b}{b\!-\!1}{b}{-3\!\lambda}00
\h\Biggr(\hs{0.2} \1by2 {\h b}{b\!-\!1}{\h b}\h -
  \Biggr(\disp{S(b\!-\!1)\over S(b)}\Biggr)^{1/2}\h
  \1by2 {\h b}{\h b}{\h b}\hs{0.3} \Biggr)& \label{12-3}
\ee
These expressions play exactly the role of  the symmetric fusion of level $2$.
In (\ref{12-1}) with $|a-b|=1$ or
(\ref{12-3}) there are two independent terms classified by
the projectors, which lead to two independent fused face weights. In
(\ref{12-1}) with $|a-b|=2$ or
(\ref{12-2}) it has only one independent term according to the independent
projector.
Thus we have the following lemma.

\begin{lemma}[Elementary Fusion] \label{lem1} If $(a,b)$ and $(d,c)$ are
admissible edges at fusion level two we define the $1\times 2$ fused
weights for shift $\lambda_1$ by
\be
\wf {W_{1 \times 2} }ubcd{\alpha}{}{\beta}{\vspace*{-0.5cm}} =
 {\disp \sum_{a'} }\;\;\phi^\alpha(a,a',b)\;
 \wt {W}a{a'}{c'}d{u} \wt {W}{a'}bc{c'}{u\+\lambda_1}
\label{eq:lem1}
\ee
\be
\phi^\alpha(a,a',b)=\left\{ \begin{array}{ll}
 (-)^{a+1-a'}\theta(a'-a)\Biggr(\disp{S(a+1)\over S(a')}\Biggr)^{1/2}
    & \mbox{if$\;a=b$, $\alpha=1$ and $\val a=3$} \\
 (-)^{a-1-a'}\theta(a-a')\Biggr(\disp{S(a-1)\over S(a')}\Biggr)^{1/2}
    & \mbox{if$\;a=b$, $\alpha=2$ and $\val a=3$} \\
(-)^{a-a'}\Biggr(\disp{S(a)\over S(a')}\Biggr)^{1/2}  & \mbox{if$\;a=b$ and
$\val a=2$} \\
\delta_{a',a+\alpha-1} & \mbox{if$\;a=b-1$} \\
\delta_{a',b+\alpha-1} & \mbox{if$\;a=b+1$} \\
\delta_{a',(a+b)/2}\delta_{\alpha,1}    & \mbox{if$\;|a-b|=2$}
\end{array} \right.
\ee
where the step function $\theta(a<0)=0$ and $\theta(a\ge 0)=1$.
The bond variable $\alpha~(\beta)=1$ for $|a-b|=2~(|c-d|=2)$ or
$a=b~(c=d)$ with $\val a~(\val c)=2$. The bond variable $\alpha=1$
and $2$ for $a=b$ with $\val a=3$ or $|a-b|=1$.
$c'\ne c$ for $c=d$ and $\val c=3$. Then it gives that
(i)
\be
\beta(c')=\left\{ \begin{array}{ll}
2 & \mbox{if $c'=c-1=d-1$ and $\val c=3$} \\
2 & \mbox{if $c'=max(c,d)$ and $|c-d|=1$}\\
1 & \mbox{otherwise.}\end{array} \right.
\ee
(ii) For all a,b,c,d the inversion relation (\ref{eq:inversion}) follows that
   $\wf {W_{_{1 \times 2}}}{0}bcd{\alpha}{\vspace*{-0.5cm}}{\beta}{}=0$.
\end{lemma}

\begin{figure}[tb]
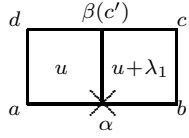

\begin{center}
\setlength{\unitlength}{0.0195in}%
\onebytwoface abcd\alpha{\beta(c')}{u}{u\!+\!\lambda_1\;}{{\sevrm .}}
\end{center}
\caption{ Elementary fusion of two faces. The cross denotes a symmetric sum
labelled by $\alpha=1,2$ as defined in lemma~1. The other spins are fixed.
If $c=d$ and $\val c=3$ we assume  that $c'\not=c$. For clarity both the
spin $c'$ and the bond variable $\beta$  are indicated.}
\end{figure}

By (\ref{oneg1p1}) and the YBR (\ref{eq:YBR}) we have
\begin{eqnarray}
\Biggr({S(c)\over S(c\+1)}\Biggr)^{1/2}\;\onebytwoface
ab{c}{c}{}{c}u{u\m+\lambda_1}{{\sevrm .}}
\hs{0.2}+  \onebytwoface ab{c}{c}{}{c\+1}u{u\m+\lambda_1}{{\sevrm .}} \hs{0.2}
   + \Biggr({S(c\-1)\over S(c\+1)}\Biggr)^{1/2}\onebytwoface
ab{c}{c}{}{c\-1}u{u
        \m+\lambda_1}{{\sevrm .}} \hs{0.2} =0.     \h \label{one1by2fu}
\ee

\noindent
Thus it follows that the fused face weights
satisfy the star-triangle relation  (\ref{eq:fuYBR}) with $l=m=1$ and $n=2$.

The antisymmetric fusion is very simple to express and is given by the
following
objector.
\be
\setlength{\unitlength}{0.0105in}%
\begin{picture}(129,117)(111,615)
\thicklines
\put(180,630){\line(-1, 1){ 30}}
\put(150,660){\line( 1, 1){ 60}}
\put(210,720){\line( 1,-1){ 30}}
\put(240,690){\line(-1,-1){ 60}}
\put(210,660){\line(-1, 1){ 60}}
\put(150,720){\line(-1,-1){ 30}}
\put(120,690){\line( 1,-1){ 30}}
\put(147,723){\scr$d$}
\put(207,723){\scr$d$}
\put(210,651){\scr$c$}
\put(180,618){\scr$b$}
\put(111,687){\scr$a$}
\put(165,657){\scr$u\!+\!\lambda_1$}
\put(147,687){\scr$u$}
\put(201,687){\scr$3\lambda$}
\end{picture}\label{antifusion}
\ee
By the property (\ref{oneg1p1}) the objector (\ref{antifusion}) gives
\begin{lemma}[Antisymmetric Fusion] \label{lem2} Put
\be
\wt {W_0}a{b}{c}d{u}={\disp \sum_{c'} }\;\;\phi(d,c',c)
 \wt {W}a{a'}{c'}d{u}\wt {W}{a'}bc{c'}{u\+\lambda_1}
             \label{eq:antilem1}
\ee
\be\phi(d,c',c)=\left\{ \begin{array}{ll}
\;\;\;0                    & \mbox{if$\;d\not=c$} \\
\Biggr(\disp{S(c')\over S(c)}\Biggr)^{1/2} & \mbox{if$\;d=c$} \\
\end{array} \right.\ee
It follows that
\be
\wt
{W_0}a{b}{c}d{u}=s_1^1(u)s^1_{-1}(u)s^1_{3/2}(u)s^1_{-3/2}(u)\delta_{a,b}\delta_{c,d}\;\; ,
\ee
where
\be
s^m_n(u)=\prod_{j=0}^{m-1}\Biggr(\disp{\vartheta_1(u+(2n-3j)\lambda)\over
              \sqrt{\vartheta_1(2\lambda)\vartheta_1(3\lambda)}}\Biggr) \; ,
\label{s}
\ee
\end{lemma}
\begin{figure}[t]
\begin{center}
\hs{-1.2}
\setlength{\unitlength}{0.0105in}%
\begin{picture}(316,210)(47,600)\thicklines
\Poperator {\lambda_1} 
\end{center}
\caption{\label{fig:P(n,-nlambdaone)}The difination of the
        $P(n,u)^{a,a_1,a_2,\cdots ,b}_{a,b_1,b_2,\cdots ,b}$.}
\end{figure}
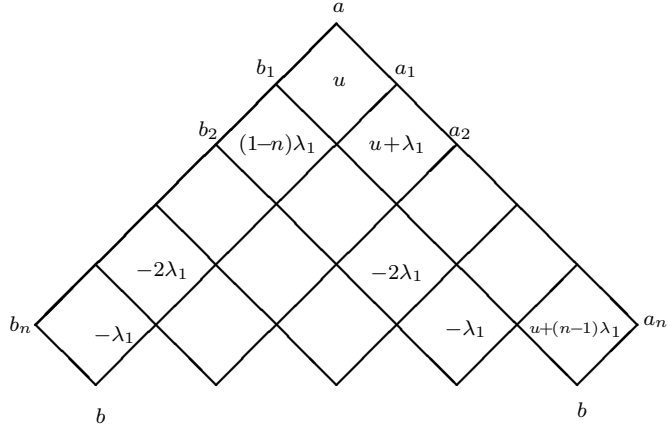

\subsection{Operator $P(n,u)$}
The operator $P(n,-n \lambda_1)$ is the projector of level $n+1$ fusion,
which element $(a,b)$ is defined graphically by
Fig.\ref{fig:P(n,-nlambdaone)} with $u=-n\lambda$. For more convenient
let us consider $P(n,u)$ defined by Fig.~$3$.
For $n=1$ it is the face weight of the elementary block and for $n=2$ it
produces the 1 by 2 fusion presented in (\ref{12-1})--(\ref{12-3}).
With the help of star-triangle relation (\ref{eq:YBR}) we can show
that the operator \vs{-2.3}satisfies
\be \hs{-0.5}
\setlength{\unitlength}{0.0115in}%
\begin{picture}(222,156)(36,726)\thicklines
\up {\lambda_1}{a_1}{a_2}{a_n}{b} \hs{-0.3}=\hs{-0.3}
\setlength{\unitlength}{0.0115in}%
\begin{picture}(222,151)(48,724)\thicklines
\down {\lambda_1}\label{eq:oneupdown}
\ee\vs{0.0}

Using (\ref{oneg2p1}) and the YBR (\ref{eq:YBR})
it is easy to see that any
two adjacent faces with the spectral parameters $u+j\lambda_1$ and
$u+(j-1)\lambda_1$ in Fig.\ref{fig:P(n,-nlambdaone)} can be considered as
the symmetric 1 by 2
fusion. So the properties (\ref{one1by2fu}) imply
\begin{eqnarray}
& & \hs{1}A_{a_{j\-1},a_{j\-1}\-1} \;
  P(n,u)_{(a,b_1,\cdots ,b_{j-1},b_j,b_{j+1},\cdots ,b)}^{
       (a,a_1,\cdots ,a_{j\-1},a_{j\-1}\-1,a_{j\+1},\cdots ,b)}
          \no \\
&&\hs{1.7} =-{S(a_{j-1})\over S(a_{j-1}\-1)}\;
 P(n,u)_{(a,b_1,\cdots ,b_{j-1},b_j,b_{j+1},\cdots ,b)}^{
    (a,a_1,\cdots ,a_{j\-1},a_{j\-1},a_{j\+1},\cdots ,b)}
    \label{eq:projprop} \\
&& \hs{2.4} -A_{a_{j\-1},a_{j\-1}\+1} {S(a_{j-1}\+1)\over S(a_{j-1}\-1)}
 P(n,u)_{(a,b_1,\cdots ,b_{j-1},b_j,b_{j+1},\cdots ,b)}^{
       (a,a_1,\cdots ,a_{j\-1},a_{j\-1}\+1,a_{j\+1},\cdots ,b)}
  \h   \for a_{j-1}=a_{j+1}    \no
\end{eqnarray}

 Let the notation $p(a,b,n)$ represent the set of all allowed paths
of $n$ steps from $a$ to $b$ on the adjacent diagrams in Figure~1 and
$P_{(a,b)}^{(n)}$ be the number of paths in the set $p(a,b,n)$.
For convenience let $p(a,b,n)_i$ represent the $i$-th path in
$p(a,b,n)$ and $p(a,b,n)_{i,j}$ be the $j$-th element of
$p(a,b,n)_i$. So we can rewrite the elements of the projector
$P(n-1,u)$ to be
  $$P(n-1,u)^{p(a,b,n)_i}_{p(a,b,n)_j}$$
The operator $P(n-1,u)$ is a suqare matrix and can be written in block
diagonal form. The block $[P(n-1,u)^{p(a,b,n)_i}_{p(a,b,n)_j}]$ is
regular because of the properties (\ref{eq:projprop}),
which give the eigenvectors of zero eigenvalues
of the block. The number of non-zero eigenvalues
of the block $[P(n-1,u)^{p(a,b,n)_i}_{p(a,b,n)_j}]$, in fact, is given
by $A^{(n)}_{(a,b)}$. Therefore the number of zero
eigenvalues of the block then is given by $P_{(a,b)}^{(n)}-A^{(n)}_{(a,b)}$.

Two paths $p(a,b,n)_i$ and  $p(a,b,n)_j$ may be related by the properties
(\ref{eq:projprop}). If so we treat the paths $p(a,b,n)_i$ and  $p(a,b,n)_j$
as dependent paths. Otherwise they are independent. Suppose there
are $m_{(a,b)}^{(n)}$ independent realtions deriving from (\ref{eq:projprop}).
So there are $A^{(n)}_{(a,b)}=P_{(a,b)}^{(n)}-
m_{(a,b)}^{(n)}$ independent paths in the set $p(a,b,n)$. We denote
these independent paths by $\alpha (a,b,n)$, $\alpha =1,2,\cdots ,
A^{(n)}_{(a,b)}$ (there may be several ways to choose the independent
paths, they give the equivalent fused models). The remaining paths
are represented in terms of the independent paths
\begin{eqnarray}
P(n-1,u)^{p(a,b,n)_i}_{\beta(a,b,n)}=\sum_{\alpha =1}^{A^{(n)}_{(a,b)}}
       \phi^{(i,\alpha)}_{(n)}(a,b) P(n-1,u)_{\beta(a,b,n)}^{\alpha(a,b,n)};
     \;\;i=1,2,\cdots ,m_{(a,b)}^{(n)}\label{eq:pathexpand}
\end{eqnarray}
for any $\beta(a,b,n)\in \{\alpha(a,b,n)|\alpha=1,2,\cdots ,
A^{(n)}_{(a,b)}\}$.
The value of $\phi^{(i,\alpha)}_{(n)}(a,b)$ takes  zero if the path
$p(a,b,n)_i$ is irrelevant to the path $\alpha(a,b,n)$ and takes
nonzero value if dependent. According to (\ref{eq:pathexpand}) we can
divide the set $p(a,b,n)$ into $A^{(n)}_{(a,b)}$ independent subsets defined by
\begin{equation}
p(n,a,\alpha,b)=\{(p(a,b,n)_i)|\phi^{(i,\alpha)}_{(n)}(a,b)\not=0\};\;
           \alpha=1,2,\cdots ,A^{(n)}_{(a,b)}.
\end{equation}
The first path in $p(n,a,\alpha,b)$ is the $\alpha(a,b,n)$, and $i$-th path is
denoted by $p(n,a,\alpha,b)_i$ and $p(n,a,\alpha,b)_{i,j}$ is the $j$-th
element
of the path $p(n,a,\alpha,b)_i$. We call $\phi^{(i,\alpha)}_{(a,b,n)}$ as the
coordinate of path $p(a,b,n)_i$ on the independent path $\alpha(a,b,n)$.
By (\ref{eq:oneupdown}) it is obvious that
\be
&& \phi^{(\alpha,\alpha)}_{(n)}(a,b)=\phi^{(i,i)}_{(n)}(a,b)=1 ,\\
&&\phi^{(i,\alpha)}_{(n)}(a,b)=\phi^{(i,\alpha)}_{(n)}(b,a) \;,\\
&&\phi^{(\beta,\alpha)}_{(n)}(a,b)=\phi^{(\alpha,\beta)}_{(n)}(a,b)=0
   \h\mbox{for $\alpha\ne\beta$}.
\ee

\subsection{General Fusion }
\label{GeneralFusionone}
Let $m$ and $n$ be positive integers. \vspace*{-0.6cm}Define
\be
& &\wf {W_{m\times n}}{u}bc{d\vs{-0.1}}{\alpha}{\nu}{\beta}{\vs{-0.1}\mu} =
{\thicklines\mnface}=\sum_{j=1}^{P^{(m)}_{(d,a)}}\!\phi^{(j,\mu)}_{(m)}(a,d)\!
 \sum_{\alpha_2,\cdots,\alpha_m} \no\\&&\no\\
& &\h \prod_{k=1}^{m}
\row {W_{1\times n}}{u\-(m\-k)\lambda_1}{p(a,d,m)_{j,k}}{\nu(b,c,m)_k}{\nu(b,
     c,m)_{k+1}}{p(a,d,m)_{j,k+1}}{\hspace*{-0.3cm}\alpha_k
        \hspace*{-0.3cm}}{\hspace*{-0.3cm}\alpha_{k+1}\hspace*{-0.3cm}}  ,
\label{eq:mnfusion}\ee
where $a=p(a,d,m)_{j,1}$, $b=\nu(b,c,m)_1$, $c=\nu(b,c,m)_{m+1}$,
$d=p(a,d,m)_{j,m+1}$, $\alpha=\alpha_1$, $\beta=\alpha_{m+1}$, the summation
$\alpha_k$'s is over $1,\cdots,A_{(p(a,d,m)_{\!_{j,k}},\nu(b,c,m)_k)}^{(n)}$.
The $1\times n$ fusion in turn is given by
\be
\wf {W_{1\times n}}{u}bcd{\alpha}{\vspace*{-0.5cm}}{\beta}{}= \hs{4}\no\\
\sum_{i=1}^{P^{(n)}_{(a,b)}}\!
\phi^{(i,\alpha)}_{(n)}(a,b)\!\prod_{k=1}^{n}  \wt
W{p(a,b,n)_{i,k}}{\hspace*{-0.3cm}p(a,b,n)_{i,k+1}}{\hspace*{-0.3cm}
 \beta(d,c,n)_{k+1}}{\beta(d,c,n)_k}{\!u\+(k\-1)\lambda_1}\;\;.
\ee

The fused face weights (\ref{eq:mnfusion}) defined here are similar to
the ones of the critical $D$  and $E$ models \cite{ZhPe:94}. In fact
the following discussion proceeds as if critical $D$ and $E$ models.
For the dilute $A_L$ models we have the following lemma:

\begin{lemma}\label{lem3}
If the path $\beta(d,c,n)$ is replaced with its dependent path
$p(n,d,\beta,c)_j$ then the fused weight
\be
\wf {W_{m\times
n}}{u}bc{d\vs{-0.1}}{\alpha}{\nu}{j}{\vs{-0.1}\mu}=\sum_{\beta'=1}^{A^{(n)}_{(d,c)}}
\phi^{(j,\beta')}_{(d,c,n)}\;
\wf {W_{m\times n}}{u}bc{d\vs{-0.1}}{\alpha}{\nu}{\beta'}{\vs{-0.1}\mu}
\label{eq:mbynfusionprop1} \ee
Similarly, if the path $\nu(b,c,m)$ is replaced by its
dependent path $p(m,b,\nu,c)_j$ then
\be
\wf {W_{m\times
n}}{u}bc{d\vs{-0.1}}{\alpha}{j}{\beta}{\vs{-0.1}\mu}=\sum_{\nu'=1}^{A^{(m)}_{(b,c)}}
\phi^{(j,\nu')}_{(b,c,m)}\;
\wf {W_{m\times n}}{u}bc{d\vs{-0.1}}{\alpha}{\nu'}{\beta}{\vs{-0.1}\mu}
\label{eq:mbynfusionprop2}.\ee
\end{lemma}

By applying YBE (\ref{eq:YBR}) to the tensor products of
$m$ by $n$ elementary blocks and with the help of
the Lemma~\ref{lem3} we can obtain the following theorem.
\begin{theorem}\label{thr1}
For a triple of positive integers $m,n,l$, the fused face weights
(\ref{eq:mnfusion}) satisfy the following star-triangle relation
\be
&&\hs{-0.5}\hspace{-0.2cm}\sum_{(\eta_1,\eta_2,\eta_3)}\!\sum_g \mbox{\small $
   \Wf fugd{e\vs{-0.2}}{\mh\eta_1\mh}{\eta_3}{\mh\mu\mh}{\nu\vs{-0.2}}
   \Wfml
g{u\-v}bc{d\vs{-0.2}}{\mh\eta_2\mh}{\beta}{\mh\gamma\mh}{\eta_3\vs{-0.2}}
   \Wfln a{v}bg{f\vs{-0.2}}{\mh\alpha\mh}{\eta_2}{\mh\eta_1\mh}{\rho\vs{-0.2}}
$}\no\\
&& \label{eq:fuYBR}\\
&&\hs{-0.5}\hspace{-0.2cm}=\!\sum_{(\eta_1,\eta_2,\eta_3)}\!\sum_g \mbox{\small
$
    \Wf aubc{g\vs{-0.2}}{\mh\alpha\mh}{\beta}{\mh\eta_1\mh}{\eta_3\vs{-0.2}}
    \Wfml f{u\-v}ag{e\vs{-0.2}}{\mh\rho\mh}{\eta_3}{\mh\eta_2\mh}{\nu\vs{-0.2}}
    \Wfln gvcd{e\vs{-0.2}}{\mh\eta_1\mh}{\gamma}{\mh\mu\mh}{\eta_2\vs{-0.2}}$}
\no\ee
\end{theorem}

By construction it is obvious that
$\wf {W_{m\times n}}{u}bc{d\vs{-0.1}}{\alpha}{\nu}{\beta}{\mu\vs{-0.1}}$
vanishes unless
\be
A^{(n)}_{a,b}&\not=&0\and \alpha=1,2,\cdots,A^{(n)}_{a,b} \no\\
A^{(n)}_{d,c}&\not=&0\and \beta=1,2,\cdots,A^{(n)}_{d,c} \no\\
A^{(m)}_{d,a}&\not=&0\and \mu=1,2,\cdots,A^{(m)}_{d,a} \no\\
A^{(m)}_{c,b}&\not=&0\and
\nu=1,2,\cdots,A^{(m)}_{c,b}  \no
\ee
where the fused adjacency matrices are given by (\ref{eq:adjfusionone}).
In particular the fused face weights exist  for any higher fusion
levels because we always have $A^{(n)}\ne 0$, $A^{(m)}\ne 0$ for $n,m>0$.

In conclusion we have constructed the fusion of the dilute $A_L$ models
for the shift $\lambda_1$. The Lemma~\ref{lem2} implies that the level $2$
antisymmetric fusion is nothing but a trivial function of the spectral
parameter and the crossing parameter.  This behavior is similar to the
fusion of the classical $A$-$D$-$E$ models. The fusion of the dilute
$A_L$ models with the shift $\lambda_1$ is therefore thought as $su(2)$
type of fusion. This is consistent with the $su(2)$ adjacency fusion rule
(\ref{eq:adjfusionone}).

\section{Fusion (1,1) in su($3$) hierarchy}
\setcounter{equation}{0}

The operators with the shift $2\lambda$ give the $su(3)$ type fusion
of the dilute $A_L$ models. The fusion procedure has been constructed
in \cite{ZPG:94}. In this section we show some connections between
the $su(2)$ fusion and  the $su(3)$ fusion.

\subsection{Adjacency conditions}
\setcounter{equation}{0}
The fusion with the shift $\lambda_2$ in structure is more interesting
than the fusion with the shift $\lambda_1$. For the $su(3)$ fusion
we need two numbers to label the fusion level. The adjacency
matrices $A^{(n,m)}$ of the level $(n,m)$ fused models are
determined by the $su(3)$ fusion rules
\begin{eqnarray}
& & A^{(n,m)} A^{(1,0)}=A^{(n+1,m)}+A^{(n-1,m+1)}+A^{(n,m-1)}\;,
                       \label{eq:adjfusion} \\
& & A^{(0,0)}={\bf I}\;,\; \;A^{(1,0)}={\bf I}+A\;,\;\; A^{(n,m)}=A^{(m,n)}
           \label{eq:adjfusion1}\\
&&A^{(n,m)}=0\;\;{\rm if} \;m,n<0\;\;{\rm or}\;\;n+m>2L-1
\label{eq:adjfusion2}\end{eqnarray}
where ${\bf I}$ is the unitary matrix. $A$ is the adjacency matrix
(\ref{ABFadj}). Like the $su(2)$ fusion, the elements
of $A^{(n,m)}$ are nonnegative integers greater than or equal to one.
The important point is that the $su(3)$ fusion rule has the
closure condition $A^{(n,m)}=0$ for $n+m=2L$. Therefore the fusion hierarchies
are truncated at the level $(m,2L-m)$ with $m=0,1,\cdots,2L$.

This $su(3)$ fusion rule is so different from previous $su(2)$ fusion
rule. However they are not totally independent. Comparing them  we find
\be
A^{(2)}=A^{(1,1)}\; .
\ee
It follows that the $su(2)$ fusion of level $2$ share the same adjacency
condition with the $su(3)$ fusion of level $(1,1)$ (see Fig.\ref{fig.3}).
In fact the fused face weights of these two fusions
are the same up to some common functions.
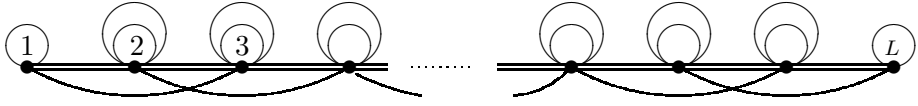
\begin{figure}[tb]
\begin{center}
\setlength{\unitlength}{0.0125in}
\begin{picture}(381,55)(0,-10)
\drawline(9.000,13.000)(13.612,10.527)(18.359,8.325)
	(23.226,6.401)(28.195,4.762)(33.252,3.414)
	(38.378,2.361)(43.556,1.606)(48.769,1.152)
	(54.000,1.000)(59.231,1.152)(64.444,1.606)
	(69.622,2.361)(74.748,3.414)(79.805,4.762)
	(84.774,6.401)(89.641,8.325)(94.388,10.527)
	(99.000,13.000)
\drawline(54.000,13.000)(58.612,10.527)(63.359,8.325)
	(68.226,6.401)(73.195,4.762)(78.252,3.414)
	(83.378,2.361)(88.556,1.606)(93.769,1.152)
	(99.000,1.000)(104.231,1.152)(109.444,1.606)
	(114.622,2.361)(119.748,3.414)(124.805,4.762)
	(129.774,6.401)(134.641,8.325)(139.388,10.527)
	(144.000,13.000)
\drawline(147.000,10.000)(152.189,7.619)(157.499,5.523)
	(162.915,3.718)(168.421,2.209)(174.000,1.000)
\drawline(282.000,13.000)(286.612,10.527)(291.359,8.325)
	(296.226,6.401)(301.195,4.762)(306.252,3.414)
	(311.378,2.361)(316.556,1.606)(321.769,1.152)
	(327.000,1.000)(332.231,1.152)(337.444,1.606)
	(342.622,2.361)(347.748,3.414)(352.805,4.762)
	(357.774,6.401)(362.641,8.325)(367.388,10.527)
	(372.000,13.000)
\drawline(237.000,13.000)(241.612,10.527)(246.359,8.325)
	(251.226,6.401)(256.195,4.762)(261.252,3.414)
	(266.378,2.361)(271.556,1.606)(276.769,1.152)
	(282.000,1.000)(287.231,1.152)(292.444,1.606)
	(297.622,2.361)(302.748,3.414)(307.805,4.762)
	(312.774,6.401)(317.641,8.325)(322.388,10.527)
	(327.000,13.000)
\drawline(213.000,1.000)(218.448,1.795)(223.700,3.444)
	(228.625,5.906)(233.096,9.118)(237.000,13.000)
\put(54,13){\circle*{6}}
\put(99,13){\circle*{6}}
\put(54,22){\circle{18}}
\put(144,13){\circle*{6}}
\put(99,22){\circle{18}}
\put(144,22){\circle{18}}
\put(9,13){\circle*{6}}
\put(9,22){\circle{18}}
\put(327,13){\circle*{6}}
\put(282,13){\circle*{6}}
\put(327,22){\circle{18}}
\put(237,13){\circle*{6}}
\put(282,22){\circle{18}}
\put(237,22){\circle{18}}
\put(54,27){\circle{26}}
\put(99,27){\circle{26}}
\put(144,27){\circle{26}}
\put(282,27){\circle{26}}
\put(327,27){\circle{26}}
\put(237,27){\circle{26}}
\put(372,22){\circle{18}}
\put(372,13){\circle*{6}}
\dottedline{3}(170,13)(194,13)
\drawline(9,14)(160,14)
\drawline(206,14)(372,14)
\drawline(9,12)(160,12)
\drawline(206,12)(372,12)
\put(6,18){\small$1$}
\put(52,18){\small$2$}
\put(96,18){\small$3$}
\put(368,18){\scr$L$}
\end{picture}
\end{center}
\caption{\label{fig.3}The graph of adjacency condition of the $su(2)$ fusion
 $A^{(2)}$ or the $su(3)$ fusion $A^{(1,1)}$. }
\end{figure}
\subsection{1 by 3 Fusion: $(1,1)$}
\setcounter{equation}{0}
The fusion of level $(1,1)$
is given by the following object

$$
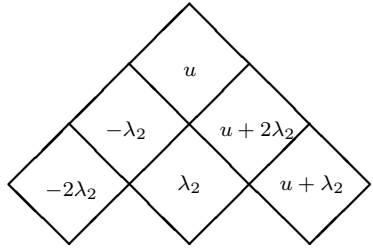
\setlength{\unitlength}{0.0105in}%
\begin{picture}(180,120)(75,660)
\thicklines
\put(165,780){\line( 1,-1){ 90}}
\put(255,690){\line(-1,-1){ 30}}
\put(225,660){\line(-1, 1){ 90}}
\put(135,750){\line(-1,-1){ 30}}
\put(105,720){\line( 1,-1){ 60}}
\put(165,660){\line( 1, 1){ 60}}
\put(195,750){\line(-1,-1){ 90}}
\put(105,660){\line(-1, 1){ 30}}
\put( 75,690){\line( 1, 1){ 30}}
\put(135,750){\line( 1, 1){ 30}}
\put(162,744){\scr $u$}
\put(159,687){\scr $\lambda_2$}
\put(210,687){\scr $u+\lambda_2$}
\put(180,714){\scr $u+2\lambda_2$}
\put( 93,684){\scr $-2\lambda_2$}
\put(123,714){\scr $-\lambda_2$}
\end{picture}$$

\noindent
which can be written as
\be
\setlength{\unitlength}{0.0105in}%
\begin{picture}(210,120)(45,645)
\thicklines
\put(135,765){\line( 1,-1){ 90}}
\put(225,675){\line(-1,-1){ 30}}
\put(195,645){\line(-1, 1){ 90}}
\put(105,735){\line( 1, 1){ 30}}
\put(165,735){\line(-1,-1){ 30}}
\put(225,735){\line( 1,-1){ 30}}
\put(255,705){\line(-1,-1){ 30}}
\put(225,735){\line(-1,-1){ 60}}
\put(135,705){\line(-1,-1){ 60}}
\put( 75,645){\line(-1, 1){ 30}}
\put( 45,675){\line( 1, 1){ 60}}
\put( 75,705){\line( 1,-1){ 30}}
\put(132,732){\scr $u$}
\put(216,699){\scr $\lambda_2$}
\put( 63,672){\scr $-2\lambda_2$}
\put( 93,702){\scr $-\lambda_2$}
\put(153,702){\scr $u+\lambda_2$}
\put(180,672){\scr $u+2\lambda_2$}
\end{picture}\label{fusion(1,1)}
\ee

\noindent
It can be seen that two projectors are used for the fusion (1,1).
The bottom one (left side) is the fully symmetric projector and the top
one (right side) is the fully antisymmetric projector, which is the elementary
face with the spectral parameter $u=\lambda_2$.  Instead of using the
projectors we introduce two fusion parities
$\psi_{(1,1)}^\alpha(d,d',c',c)$ on the top and
$\phi_{(1,1)}^\alpha(a,a',b',b)$ on the bottom, which can be given
by selecting the independent paths of these projectors. However,
more simpler way is to do the antisymmetric fusion first on the top.
It has been known that the antisymmetric fused weights are
equivalent to the elementary face weights in following way
\cite{ZPG:94},
\be
\setlength{\unitlength}{0.0125in}%
\begin{picture}(52,78)(221,696)
\thicklines
\put(150,735){\circle*{4}}
\put(120,735){\line( 1, 0){ 30}}
\put(120,765){\line( 0,-1){ 60}}
\put(120,705){\line( 1, 0){ 30}}
\put(150,705){\line( 0, 1){ 60}}
\put(150,765){\line(-1, 0){ 30}}
\put(172,756){\line(-1,-1){ 21.500}}
\put(150,735){\line( 1,-1){ 22}}
\put(172,713){\line( 1, 1){ 21.500}}
\put(193,735){\line(-1, 1){ 21}}
\put(132,744){\scr$u$}
\put(124,718){\scr$u\!-\!2\lambda$}
\put(167,731){\scr$2\lambda$}
\put(114,696){\scr$a$}
\put(111,765){\scr$b$}
\put(171,759){\scr$c$}
\put(171,702){\scr$d$}
\put(153,768){\scr$c$}\put(210,730){\mbox{$\sim$$\;\;-s^1_0(u)s^1_{-5/2}(u)$}}
\put(153,696){\scr$d$}
\end{picture}
\setlength{\unitlength}{0.0125in}%
\begin{picture}(46,56)(50,717)
\thicklines
\put(181,777){\line( 1,-1){ 21}}
\put(202,756){\line(-1,-1){ 21.500}}
\put(181,734){\line(-1, 1){ 22}}
\put(159,756){\line( 1, 1){ 21.500}}
\put(180,720){\scr$d$}
\put(180,780){\scr$c$}
\put(150,753){\scr$c'$}
\put(175,751){\scr$2\lambda$}
\put(105,768){\line( 1, 0){ 30}}
\put(135,768){\line( 0,-1){ 30}}
\put(135,738){\line(-1, 0){ 30}}
\put(105,738){\line( 0, 1){ 30}}
\put( 96,729){\scr$a$}
\put(138,771){\scr$c$}
\put( 96,768){\scr$b$}
\put(138,729){\scr$d$}
\put(111,750){\scr$u\!-\!\lambda$}
\end{picture}
\ee
\be
\setlength{\unitlength}{0.0125in}%
\begin{picture}(42,65)(175,726)
\thicklines
\put(117,765){\circle*{4}}
\put(117,735){\line( 0, 1){ 30}}
\put( 87,735){\line( 1, 0){ 60}}
\put(147,735){\line( 0, 1){ 30}}
\put(147,765){\line(-1, 0){ 60}}
\put( 87,765){\line( 0,-1){ 30}}
\put( 96,787){\line( 1,-1){ 21.500}}
\put(117,765){\line( 1, 1){ 22}}
\put(139,787){\line(-1, 1){ 21.500}}
\put(117,808){\line(-1,-1){ 21}}
\put(150,765){\scr$c$}
\put(150,726){\scr$b$}
\put( 78,765){\scr$d$}
\put( 81,726){\scr$a$}
\put( 87,780){\scr$d$}
\put(144,780){\scr$c$}
\put( 96,747){\scr$u$}
\put(121,746){\scr$u\!+\!2\lambda$}
\put(112,782){\scr$2\lambda$}\put(170,750){\mbox{$\sim$$\;\;-s^1_1(u)s^1_{-3/2}(u)$}}
\end{picture}
\setlength{\unitlength}{0.0125in}%
\begin{picture}(21,76)(30,721)
\thicklines
\put(108,717){\line( 0, 1){ 30}}
\put(108,747){\line( 1, 0){ 30}}
\put(138,747){\line( 0,-1){ 30}}
\put(138,717){\line(-1, 0){ 30}}
\put( 99,793){\line( 1, 1){ 21}}
\put(120,814){\line( 1,-1){ 21.500}}
\put(142,793){\line(-1,-1){ 22}}
\put(120,771){\line(-1, 1){ 21.500}}
\put(117,759){\scr$c'$}
\put(144,744){\scr$c$}
\put( 99,744){\scr$d$}
\put(144,711){\scr$b$}
\put(111,729){\scr$u+\lambda$}
\put(115,788){\scr$2\lambda$}
\put( 93,786){\scr$d$}
\put(144,786){\scr$c$}
\put( 99,711){\scr$a$}
\end{picture} \label{spe}
\ee
Where $c'=min(c,d)$ for $|c-d|=1$ and $c'=c$ for $|c-d|=0$. In (\ref{spe})
the projector graphically is the square face with anti-clockwise rotation
through $\pi\over 2$.
By these relations the objector (\ref{fusion(1,1)}) can be simply reduced to
(\ref{12fusion}).
So we have the following lemma.

\begin{lemma}\label{lem7} Put
$$\wf
{W^{(1,1)}_{(1,0)}}ubcd{\alpha}{}{\beta}{\vspace*{-0.5cm}}={\disp\sum_{a',b',c',d'}}
   \psi_{(1,1)}^\beta(d,d',c',c)\;\phi_{(1,1)}^\alpha(a,a',b',b)  $$
\be
\h\times\wt Wa{a'}{d'}d{u} \wt W{a'}{b'}{c'}{d'}{u\+\lambda_2}
    \wt W{b'}bc{c'}{u\+2\lambda_2}. \label{eq:oneonefu} \ee
Then it follows that

(i) This fusion is equivalent to the $su(2)$ fusion of level $2$. Or
disregarding the trivial gauge factors we have
\be
\wf
{W^{(1,1)}_{(1,0)}}ubcd{\alpha}{}{\beta}{\vspace*{-0.5cm}}=-s^1_2(u)s^1_{-1/2}(u)
\wf {W_{_{1 \times 2}}}ubcd{\alpha}{}{\beta}{\vspace*{-0.5cm}}
\ee
(ii) For all fixed $a,b,c,d,\alpha,\beta$ we have
     $$\wf {W^{(1,1)}_{(1,0)}}{u}bcd{\alpha}{\vspace*{-0.5cm}}{\beta}{}=0
\h \mbox{if $u=0,\lambda,-2\lambda_2$}.$$
\end{lemma}

\section{Su($2$) Fusion hierarchy}
\label{Fusionhierarchy}
\setcounter{equation}{0}

The fusion rule (\ref{eq:adjfusionone}) is the relations for the
adjacency matrices of the  fused models. We will see in this section
that the theory carries over to the level of the row transfer matrix.

Suppose that ${\bf a}$ (\mbox{\boldmath $\alpha$}) and ${\bf b}$
(\mbox{\boldmath $\beta$}) are allowed spin (bond) configurations
of two consecutive rows of an $N$ (even) column lattice with periodic
boundary conditions. The elements of the fused row transfer matrix
${\bf T}(u)$ are given \vspace*{-1.1cm}by
\begin{eqnarray}
\langle\mbox{\boldmath $a,\alpha$}|{\bf T}_{(m,n)}(u)|
       \mbox{\boldmath $b,\beta$}\rangle =
\prod_{j=1}^N\!\sum_{\eta_j} \Wf
{a_j}u{b_j}{b_{j\+1}}{a_{j\+1}\vs{-0.1}}{\mh\eta_j
        \mh}{\beta_j}{\mh\eta_{j\+1}\mh}{\alpha_j\vs{-0.1}}=
\setlength{\unitlength}{0.0115in}%
\begin{picture}(48,90)(60,760)
\thicklines
\put( 75,780){\line( 0,-1){ 36}}
\put( 75,744){\line( 1, 0){ 30}}
\put(105,744){\line( 0, 1){ 36}}
\put(105,780){\line(-1, 0){ 30}}
\multiput(105,810)(0,-7.82609){12}{\line(0,-1){3.913}}
\multiput( 75,810)(0,-8.28571){11}{\line(0,-1){4.143}}
\put( 87,759){\ma{\ra{\twlrm\scr $u$}}}
\put( 64,756){\ma{\ra{\twlrm \scr $\alpha_j$}}}
\put(108,756){\ma{\ra{\twlrm \scr $\beta_j$}}}
\put(108,774){\ma{\ra{\twlrm \scr $b_{j\+1}$}}}
\put(108,738){\ma{\ra{\twlrm \scr $b_j$}}}
\put( 64,738){\ma{\ra{\twlrm \scr $a_j$}}}
\put( 60,774){\ma{\ra{\twlrm \scr $a_{\!j\+1}$}}}
\end{picture}
\ee\sk{0.3}
where $a_{N+1}=a_1$, $b_{N+1}=b_1$ and $\eta_{N+1}=\eta_1$. Specifically,
the star-triangle relation (\ref{eq:fuYBR}) of the fused weights
(\ref{eq:mnfusion}) imply the commutation relations
\begin{equation}
[{\bf T}_{(m,n)}(u),{\bf T}_{(m,\ol{n})}(v)] = 0.
\label{eq:rowcommute1}
\end{equation}
Thus if fix $m$ we have the hierarchy of commuting families of
transfer matrices. The fusion procedure implies various relations among these
transfer matrices. We summarize them in following theorems.

\begin{theorem}[$su(2)$ Fusion Hierarchy]
Let us define
\be
\T^{(n)}_k&=&{\T}_{(m,n)}(u+3k\lambda)           \no \\
\T^{(n)}&=&0   \h \mbox{if $n<0$ or $m<0$}       \no \\
\T^{(0)}&=&{\I}                                  \no \\
 f^m_n&=&
\left[
s^m_{3n/2+1}(u)s^m_{3n/2-1}(u)s^m_{3(n+1)/2}(u)s^m_{3(n-1)/2}(u)\right]^N
\ee
Then for $n\ge 0$
\be
&&\hs{2}\T^{(n)}_0\T^{(1)}_n= \T^{(n+1)}_0  + f^m_{n-1}\T^{(n-1)}_0
\label{fhT4}
\ee
without closure condition.
\end{theorem}

Starting with the fusion hierarchy we can easily derive new functional
equations
\be
\T^{(n)}_0\T^{(n)}_1=\I\prod_{k=0}^{n-1}f^m_k  + \T^{(n+1)}_0\T^{(n-1)}_1
\label{TT}
\ee
Then it is easy to see the following theorem.
\begin{theorem}[$su(2)$ TBA] If we define
\be
\t^0_0&=&0 \;     \\
\t^n_0&=&\disp{\T^{(n+1)}_0\T^{(n-1)}_1\over\I \prod_{k=0}^{n-1}f^m_k }\; ,
\ee
then it follows that $su(2)$ TBA-like equations
\be
\t^n_0\t^n_1=(\I+\t^{n+1}_0)(\I+\t^{n-1}_1)
\ee
without closure condition.
\end{theorem}

The fusion hierarchy (\ref{fhT4}) and the fusion rule (\ref{eq:adjfusionone})
are similar in form. They can be associated with affine $su(2)$ and there is
an one-to-one correspondence between $\T^{(n)}$ or $A^{(n)}$ and a Young
diagram. Let us represent $\T^{(n)}$ or $A^{(n)}$ by a Young diagram
with $n$ blocks in one row. Identify any Young diagram with two rows
as an one row Young diagram by subtracting the columns of length 2.
Then the fusion hierarchy (\ref{fhT4}) or the fusion rule
(\ref{eq:adjfusionone})
can be represented graphically by
\def\A{\mbox{$\tilde A$}}
\def\Aoneo{\setlength{\unitlength}{0.008in}%
\begin{picture}(70,40)(-5,10)
\multiput(0,0)(30,0){3}{\line(0,1){30}}
\multiput(0,0)(0,30){2}{\line(1,0){60}}
\end{picture}}
\def\Atwon#1#2#3#4{\setlength{\unitlength}{0.008in}%
\begin{picture}(#1,45)(0,0)
\multiput(0,0)(#1,0){2}{\line(0,1){30}}
\multiput(0,0)(0,30){2}{\line(1,0){#1}}
\put(#1,6){$\hs{0.3}#2$}\put(#3,-19){\scr$#4$}
\end{picture}}
\def\Aone{\hs{0.85}\Atwon {50}{}{}{}}
\be
&&\underbrace{\Atwon {180}{\otimes}{85}{n}} \hs{0.85}{\Atwon {30}{}{5}{}} \no
\\
&&\;=\;\underbrace{\Atwon {150}{\oplus}{55}{n-1}}\hs{0.85}
 \underbrace{\Atwon {210}{}{85}{n+1}}
\ee
and the functional relations (\ref{AA}) or (\ref{TT}) are represented by
\be
&&\underbrace{\Atwon {180}{\otimes}{85}{n}} \hs{0.85}\underbrace{\Atwon
{180}{}{85}{n}} \no \\
&&\;=\;\underbrace{\Atwon {150}{\otimes}{55}{n-1}}\hs{0.85}
 \underbrace{\Atwon {210}{\oplus\hs{0.2}{\sc\bullet}}{85}{n+1}} \hs{0.85}
\ee
where the dot means the identity matrix.
\section{Bethe Ansatz}

The eigenvalues $\Lambda_{(1,0)}(u)$ and the Bethe ansatz equations of the row
transfer matrices $\T(u)$ has been given in
\cite{BNW:94}. Obviously, the eigenvalues and the Bethe ansatz
equations can be extended to the transfer matrix $\T_{(m,1)}(u)$ and they are
given by
\be
\Lambda_{(m,1)}(u) & = &
\omega\,[s^m_{-1}(u)s^m_{-3/2}(u)]^N\frac{Q^{(m)}(u+\lambda)}{Q^{(m)}(u-\lambda)}
      + \omega^{-1}\,[s^m_0(u)s^m_{-1/2}(u)]^N
           \frac{Q^{(m)}(u-4\lambda)}{Q^{(m)}(u-2\lambda)}  \no\\*
     &   & \qquad + [(-1)^ms^m_0(u)s^m_{-3/2}(u)]^N
\frac{Q^{(m)}(u)Q^{(m)}(u-3\lambda)}{Q^{(m)}(u-\lambda)Q^{(m)}(u-2\lambda)}
\label{BAE}
\ee
where
\be
Q^{(m)}(u)=\prod_{j=1}^{mN} \vartheta_1(u-iu_j)
\ee
and the zeros $\{u_j\}$ satisfy the Bethe ansatz equations
\be
\omega \Biggr[\frac{s^m_{1/2}(iu_j)}{s^m_{-1/2}(iu_j)}\Biggr]^N \!\!& = &
(-1)^{mN+1} \prod_{k=1}^{mN}
\frac{\vartheta_1(iu_j-iu_k+2\lambda)\vartheta_1(iu_j-iu_k-\lambda)}
{\vartheta_1(iu_j-iu_k-2\lambda)\vartheta_1(iu_j-iu_k+\lambda)}
\ee
with $j=1,\cdots,N$ and $\omega=\exp(i\pi\ell/(L+1))$, $\ell=1,\cdots,L$.
The Bethe ansatz equations ensure that the eigenvalues $\Lambda_{(m,1)}(u)$
are entire functions of $u$. Inserting the solution $\Lambda_{(m,1)}(u)$
into the functional relations (\ref{fhT4}) we can obtain the eigenvalues
of the transfer matrices of whole hierarchy. In particular, the eigenvalues
of $\T_{(2,2)}$ are given by
\be
\Lambda_{(2,2)}(u)
&=&\Lambda_{(2,1)}(u)\Biggr(
\omega\,[s^2_{0}(u)s^2_{1/2}(u)]^N{Q^{(2)}(u+4\lambda)
                                   \over Q^{(2)}(u+2\lambda)}  \no \\ \no \\
&&\hs{0.7} +\;[s^2_{3/2}(u)s^2_0(u)]^N{Q^{(2)}(u+3\lambda)Q^{(2)}(u)
                      \over Q^{(2)}(u+2\lambda)Q^{(2)}(u+\lambda)} \Biggr)
\no\\ \no\\
&&\;+\;\omega^{-1}\,[s^2_1(u)s^2_{3/2}(u)]^N\frac{Q^{(2)}(u-\lambda)}
                                                 {Q^{(2)}(u+\lambda)} \no \\
\no \\
&&\h
\times\;\Biggr(\omega^{-1}\,[s^2_0(u)s^2_{-1/2}(u)]^N\frac{Q^{(2)}(u-4\lambda)}
                                                     {Q^{(2)}(u-2\lambda)}\no\\
\no\\
&&\hs{0.7}+\;[s^2_0(u)s^2_{-3/2}(u)]^N\frac{Q^{(2)}(u)Q^{(2)}(u-3\lambda)}
                     {Q^{(2)}(u-\lambda)Q^{(2)}(u-2\lambda)}\Biggr)
\ee
\smallskip
This represents also the eigenvalues of the transfer matrix
\begin{eqnarray}
\langle\mbox{\boldmath $a,\alpha$}|{\bf T}^{(1,1)}_{(1,1)}(u)|
       \mbox{\boldmath $b,\beta$}\rangle =
\prod_{j=1}^N\!\sum_{\eta_j} \w11 {a_j}u{b_j}{b_{j\+1}}{a_{j\+1}}{\mh\eta_j
        \mh}{\beta_j}{\mh\eta_{j\+1}\mh}{\alpha_j}
\ee
for the fusion level $(1,1)\times(1,1)$ in $su(3)$ fusion hierarchy.

\section{Discussion}

We have presented the $su(2)$ fusion procedure for the dilute $A_L$ models.
The fusion of critical dilute $D$ and $E$ models can be implemented in a
similar way \cite{Zhou:95}.
The functional relations of the $su(2)$ fusion hierarchy do
not close and the fusion exists for any higher level. This likes
the fusion hierarchy of the vertex models. However the dilute models are
the restricted SOS models. We have infinite number of the fused models
in this $su(2)$ fusion hierarchy and all these fused models are the restricted
SOS models. This behavior is generally not common for the restricted
SOS models. So we need to study further to understand this fusion
hierarchy. Particularly, it is interesting to find the finite-size corrections
of
the transfer matrices of the $su(2)$ fused models. In the $su(3)$ hierarchy the
finite-size corrections of the dilute models have been studied \cite{ZhPe:94b}.

As a possible way we may treat the $su(2)$ fusion hierarchy like
the spin $1$ representation of $su(2)$. This however is allowed for
the adjacency matrices. Attach the Young diagrams to the adjacency
matrix $A^{(1)}$ and $A^{(0)}=\I$ in following ways,
\be
\A^{(2)}:=A^{(1)}\h\sim \h\Aoneo
  \;=\begin{picture}(50,0)(0,0)\put(-28,-10){\Aone}\end{picture}
\ee
\be
\A^{(0)}:=A^{(0)}\h\sim \h
\setlength{\unitlength}{0.008in}%
\begin{picture}(40,70)(-5,20)
\multiput(0,0)(30,0){2}{\line(0,1){60}}
\multiput(0,0)(0,30){3}{\line(1,0){30}}
\put(60,18){$=$\scr$\h\bullet$}
\end{picture}
\ee

\noindent
We can have the following $su(2)$ fusion rule
\be
\A^{(2n)}\A^{(2)}=\A^{(2n-2)}+\A^{(2n)}+\A^{(2n+2)} \label{newsu(2)}
\ee
with
\be
\A^{(2L)}=Y \; ,
\ee
where $n=1,2,\cdots,L-1$ and $Y$ is the spin reversal operator or
$Y_{a,b}=\delta_{a,L-b}$.
This fusion rule is of $su(2)$ type, but, differs to the $su(2)$
fusion rule (\ref{eq:adjfusionone}). We may derive the new fusion
rule (\ref{newsu(2)}) with Young diagrammatic methods as follows.
\be
&&\underbrace{\Atwon {180}{\otimes}{82}{2n}}\Aone \no \\
&&\;=\;\underbrace{\Atwon {150}{\oplus}{52}{2n-2}}\hs{0.85}
   \underbrace{\Atwon {180}{\oplus}{82}{2n}} \hs{0.85}  \no \\
&&\h\; \underbrace{\Atwon {210}{}{82}{2n+2}}
\ee
where we have used the graphical representation
\be
\A^{(2n)}\h\sim\h \underbrace{\Atwon {180}{.}{82}{2n}}
\ee

This new fusion rule shares the $\Z_2$ symmetry
\be
\A^{(2L-2n)}=Y\A^{(2n)}=\A^{(2n)}Y\; ,\h n=0,1,\cdots,L\; .
\ee
Therefore it is easy to see that
at each level $2kL+1$ for any positive integer $k$ it starts to repeat
the fusion rule (\ref{newsu(2)}) again
and thus this fusion rule is truncated. Here it has been shown that
this procedure works for the adjacency matrices. It is not clear how it
works for the transfer matrices.

\section*{Acknowledgements}
The author thanks Paul Pearce and Ole Warnaar for discussion.
This research is supported by the Australian Research Council.

\clearpage
\typeout{}
\typeout{}
\end{document}